\documentclass[prd,draft,eqsecnum,showpacs,aps,nofootinbib]{revtex4}

\usepackage{latexsym}
\usepackage[dvips]{graphicx} 

\newcommand{\beq}{\begin{equation}}
\newcommand{\eeq}{\end{equation}}
\newcommand{\beqarr}{\begin{eqnarray}}
\newcommand{\eeqarr}{\end{eqnarray}}

\def\me{m_e}
\def\te{T_e}

\def\cos{{\rm cos}}

\def\bp{ {\bf p} }
\def\bpp{ {\bf p'} }

\def\bv{ {\bf v} }

\def\fz{f^{(0)}(p)}
\def\fpz{f^{(0)}(p')}
\def\f#1{f^{(#1)}(\bp)}
\def\fps#1{f^{(#1)}(\bpp)}

\def\part#1;#2 {\partial#1 \over \partial#2}
\def\deriv#1;#2 {d#1 \over d#2}

\def\bea#1#2{\global\advance\counteqn by 1
\xdef#1{(\number\countsec.\number\counteqn)}
$$\eqalignno{#2 &(\number\countsec.\number\counteqn)\cr}$$}

\begin{document}
\draft

%
%
%
\renewcommand{\topfraction}{0.99}
\renewcommand{\bottomfraction}{0.99}
\title{CMB Anisotropies at Second-Order I 

}

\author{Nicola Bartolo}\email{nbartolo@ictp.trieste.it}
\affiliation{The Abdus Salam International Centre for Theoretical 
Physics, Strada Costiera 11, 34100 Trieste, Italy}

\author{Sabino Matarrese}\email{sabino.matarrese@pd.infn.it}
\affiliation{Dipartimento di Fisica ``G.\ Galilei'', Universit\`{a} di Padova, 
        and INFN, Sezione di Padova, via Marzolo 8, Padova I-35131, Italy}

\author{Antonio Riotto}\email{antonio.riotto@pd.infn.it}
\affiliation{CERN, Theory Division, CH-1211 Geneva 23, Switzerland}

\date{\today}
\vskip 2cm
\begin{abstract}
\noindent 
We present 
the computation of the full system of   
Boltzmann equations at second-order
describing the evolution of the photon, baryon and cold dark matter  fluids. 
These equations
allow to follow the time evolution of the Cosmic Microwave
Background (CMB)  anisotropies at second-order
at all angular scales
from the early epoch,  when the cosmological perturbations were generated,
to the present  through the recombination era. This
paper sets the stage for the computation of the full second-order
radiation transfer function at all scales and for a 
a generic set of initial conditions 
specifying the level of primordial non-Gaussianity. In a companion paper, 
we will present the computation of the three-point correlation
function at recombination which is so relevant for the issue of non-Gaussianity
in the CMB anisotropies.
\end{abstract}

\pacs{98.80.Cq \hfill DFPD 06/A/05
}
\maketitle
\vskip2pc
\section{Introduction}
\noindent 
Cosmological inflation 
\cite{lrreview} has become the dominant paradigm to 
understand the initial conditions for the CMB anisotropies
and structure formation. In the inflationary picture, 
the primordial cosmological perturbations are created from quantum 
fluctuations ``redshifted'' out of the horizon during an early period of 
accelerated expansion of the universe, where they remain ``frozen''.  
They are observable as temperature anisotropies in the CMB. 
This picture has recently received further   spectacular confirmations 
by the  Wilkinson Microwave 
Anisotropy Probe (WMAP) three year set of data \cite{wmap3}.
Since the observed cosmological perturbations are of the order
of $10^{-5}$, one might think that first-order perturbation theory
will be adequate for all comparisons with observations. This might not be
the case, though.  Present \cite{wmap3} and future experiments \cite{planck}  
may be sensitive to the non-linearities of the cosmological
perturbations at the level of second- or higher-order perturbation theory.
The detection of these non-linearities through the  non-Gaussianity
(NG)   in the CMB \cite{review}
has become one of the
primary experimental targets. 

There is one fundamental  reason why a positive detection of NG is so
relevant: it might help in discriminating among the various mechanisms
for the generation of the cosmological perturbations. Indeed,
various models of inflation, firmly rooted in modern 
particle physics theory,   predict a significant amount of primordial
NG generated either during or immediately after inflation when the
comoving curvature perturbation becomes constant on super-horizon scales
\cite{review}. While single-field  \cite{noi}
and two(multi)-field \cite{two} models of inflation predict
a tiny level of NG, 
`curvaton-type models',  in which
a  significant contribution to the curvature perturbation is generated after
the end of slow-roll inflation by the perturbation in a field which has
a negligible effect on inflation, may predict a high level of NG \cite{luw}.
Alternatives to the curvaton model are those models 
characterized by the curvature perturbation being 
generated by an inhomogeneity in 
the decay rate \cite{hk,varcoupling},
  the mass   \cite{varmass} or 
the interaction rate \cite{thermalization}
of the particles responsible for the reheating after inflation. 
In that case the 
reheating can be the first one (caused by the scalar field(s) responsible
for the energy density during inflation) or alternatively the particle
species causing the reheating can be a fermion  \cite{bc}.
Other opportunities for generating the curvature perturbation occur
 at the end of inflation \cite{endinflation}, during
preheating \cite{preheating}, and 
 at a phase transition producing cosmic strings \cite{matsuda}.

Statistics like the bispectrum and the trispectrum of the 
CMB can then be used to assess the level of primordial NG on 
various cosmological scales 
and to discriminate it from the one induced by 
secondary anisotropies and systematic effects \cite{review,hu,dt,jul}. 
A positive detection of a primordial NG in the CMB at some level 
might therefore 
confirm and/or rule out a whole class of mechanisms by which the cosmological
perturbations have been generated.

Despite the importance of NG in the CMB anisotropies, little effort
has ben made so far to provide accurate
theoretical predictions of it. On the contrary, the vast majority of the
literature has been devoted to the computation of the bispectrum
of either the comovig curvature perturbation or the gravitational 
potential on large scales within given inflationary models. 
These, however, are not the  physical
quantities which are observed. One should instead provide a full prediction
for the second-order radiation transfer function. 
A firt step towards this goal has been taken in Ref. \cite{fulllarge}
(see also \cite{two})
where the full second-order radiation transfer function 
for the CMB  anisotropies on large angular scales in a flat universe 
filled with matter and cosmological constant was computed, including 
the second-order generalization of the Sachs-Wolfe effect,
both the early and late Integrated Sachs-Wolfe (ISW) effects and the 
contribution of the second-order tensor modes.

There are many sources of NG in the CMB anisotropies, beyond the primordial
one.  The most relevant  sources are the so-called  secondary anisotropies,
which arise  after the last scattering epoch. These anisotropies can be 
divided into two categories: scattering secondaries, when the CMB photons 
scatter with electrons along the line of sight, 
and gravitational secondaries when effects are mediated by 
gravity \cite{Hu:2001bc}.  Among the  scattering secondaries 
we may list the thermal Sunyaev-Zeldovich effect, where 
hot electrons in clusters transfer energy to the CMB photons, 
the kinetic Sunyaev-Zeldovich effect produced by the bulk motion of the 
electrons in clusters, the Ostriker-Vishniac effect, produced by bulk motions modulated 
by linear density perturbations, and effects due to  
reionization processes. The scattering secondaries 
are most significant on small angular scales
as density inhomogeneities, bulk and thermal motions  grow and become 
sizeable  on small length scales when 
structure formation proceeds.

Gravitational secondaries arise from the change 
in energy of photons when the gravitational potential is time-dependent, the 
ISW effect, and  gravitational lensing.  At late times, when the 
Universe becomes dominated by the dark energy, 
the gravitational potential on linear scales starts to decay, causing 
the ISW effect mainly on large angular scales. Other secondaries that result from a 
time dependent potential are the Rees-Sciama effect, produced during
the matter-dominated epoch at second-order and by the time 
evolution of the potential on non-linear scales. 

The fact that the potential never grows appreciably means that most 
second order effects created by gravitational secondaries are generically small
compared to those created by 
scattering ones. However,  when a photon propagates from the last scattering to 
us, their path may be deflected because of the gravitational lensing. This
effect 
does not create anisotropies, but it only modifies existing ones. Since
photons with large wavenumbers $k$  are lensed over many regions ($\sim k/H$, 
where $H$ is the Hubble rate) along the
line of sight, the corresponding second-order effect may be sizeable.
The  
three-point function arising from the correlation of the gravitational lensing 
effect and the ISW effect generated by  
the matter distribution along the line of sight 
\cite{Seljak:1998nu,Goldberg:xm} and the Sunyaev-Zeldovich effect \cite{sk}
are large and detectable by Planck~\cite{ks}.

Another relevant source of NG comes from the physics operating at
the recombination. A naive estimate leads to think that these non-linearities
are tiny being suppressed by an extra power of the gravitational 
potential. However, the dynamics  at recombination is quite involved because
all the non-linearities in the evolution of
the baryon-photon fluid at recombination and  the ones 
coming from general relativity should be accounted for. 
This complicated dynamics might lead
to  unexpected suppressions or enhancements of the NG at recombination. 
A step towards the evaluation of the three-point correlation function
has been taken on Ref. \cite{rec} where it was computed in the 
in so-called  squeezed triangle limit, when one mode has a 
wavelength much larger than the other two and is outside the horizon.

In this paper we present 
the computation of the full system of  
Boltzmann equations at second-order
describing the evolution of the photon, baryon and CDM  fluids, 
neglecting polarization. 
These equations
allow to follow the time evolution of the CMB anisotropies at second-order
at all angular scales
from the early epoch,  when the cosmological perturbations were generated,
to the present  through the recombination era. This
paper sets therefore the stage for the computation of the full second-order
radiation transfer function at all scales and for a 
a generic set of initial conditions 
specifying the level of primordial non-Gaussianity. In a companion paper
\cite{toappear},
we will present the computation of the three-point correlation
function at recombination which is so relevant for the issue of NG
in the CMB anisotropies. Of course on small angular scales, fully non-linear
calculations of specific effects like Sunyaev-Zel'dovich,
gravitational lensing, etc.  would provide a more
accurate estimate of the resulting CMB anisotropy, however,
as long as the leading contribution to 
second-order statistics like the bispectrum  is
concerned, second-order perturbation theory suffices.

The paper is organized as follows. In Section II we provide the 
second-order metric and corresponding Einstein equations. In Section III
the left-hand-side of the Boltzmann equation 
for the photon distribution function 
is derived at second-order. The collision term is computed in Section IV. 
In Section V, we present the second-order
Boltzmann equation for the photon brightness function, its formal
solution with the method of the integration along the line of sight and the
corresponding hierarchy equations for the multipole moments.
Sections VI contains the derivation of the Boltzmann equations at second-order
for baryons and Cold Dark Matter (CDM).  Section VII contains the expressions
for the energy momentum tensors and, finally, Section VIII
contains
our summary.

In performing the computation presented in this paper, we have mainly followed 
Ref. \cite{Dodelsonbook} (in particular chapter 4) where an excellent
derivation of the Boltzmann equations for the baryon-photon fluid
at first-order is given. Since the derivation at second-order  is 
straightforward, but lenghty, the reader might benefit 
from reading the appropriate sections of \cite{Dodelsonbook}. At the end of the paper we have 
provided a Table which summarizes the many symbols adopted appearing throughout the paper.

\section{Perturbing gravity}
\noindent 
Before tackling the problem of interest -- the computation of the Boltzmann
equations for the baryon-photon and CDM fluids -- we first provide the
necessary tools to deal with perturbed gravity, giving 
 the expressions for the Einstein tensor 
perturbed up to second-order around  a flat 
Friedmann-Robertson-Walker background, and the relevant Einstein equations. 
In the following we will adopt the Poisson gauge which eliminates one scalar degree of freedom from the $g_{0i}$ component of the 
metric and one scalar and two vector degrees of freedom from $g_{ij}$. We will use a metric of the form   
\begin{equation}
\label{metric}
ds^2=a^2(\eta)\left[
-e^{2\Phi} d\eta^2+2\omega_i dx^i d\eta+(e^{-2\Psi}\delta_{ij}+\chi_{ij}) dx^i dx^j
\right]\, ,
\end{equation}
where $a(\eta)$ is the scale factor as a function of the conformal time $\eta$, and $\omega_i$ and $\chi_{ij}$ 
are the vector and tensor peturbation modes 
respectively. Each metric perturbation can be expanded into a 
linear (first-order) and a second-order part, as for example, the gravitational potential $\Phi=\Phi^{(1)}+\Phi^{(2)}/2$. However 
in the metric~(\ref{metric}) the choice of the exponentials greatly helps in computing the relevant expressions, and thus we will 
always keep them where it is convenient. From Eq.~(\ref{metric}) one recovers at linear order the well-known  
longitudinal gauge while at second-order, one finds 
$\Phi^{(2)}=\phi^{(2)}-2 (\phi^{(1)})^2$ and $\Psi^{(2)}=
\psi^{(2)}+2(\psi^{(1)})^2$ where $\phi^{(1)}$, $\psi^{(1)}$ 
and $\phi^{(2)}$, $\psi^{(2)}$ (with 
$\phi^{(1)}=\Phi^{(1)}$ and $\psi^{(1)}=\Psi^{(1)}$) are the first and second-order gravitational 
potentials in the longitudinal (Poisson) gauge adopted in Refs.~\cite{MMB,review} as far as  scalar perturbations are concerned.
For the vector and tensor perturbations,  we will neglect linear vector modes since they are not produced in standard 
mechanisms for the generation of cosmological perturbations (as inflation), 
and we also neglect tensor modes at linear order, since they give a negligible contribution to second-order 
perturbations. Therefore we take $\omega_i$ and $\chi_{ij}$ to be 
second-order vector and tensor perturbations of the metric. 

Let us now give our definitions for the connection coefficients and their expressions for the metric~(\ref{metric}). 
The number of spatial dimensions is $n=3$.
Greek indices ($\alpha, \beta, ..., \mu, \nu, ....$)
 run from 0 to 3, while latin indices ($a,b,...,i,j,k,....
m,n...$) run from 1 to 3. 
The total spacetime metric $g_{\mu \nu}$ has signature ($-,+,+,+$). 
The connection coefficients are defined as
\begin{equation}
\label{conness} \Gamma^\alpha_{\beta\gamma}\,=\,
\frac{1}{2}\,g^{\alpha\rho}\left( \frac{\partial
g_{\rho\gamma}}{\partial x^{\beta}} \,+\, \frac{\partial
g_{\beta\rho}}{\partial x^{\gamma}} \,-\, \frac{\partial
g_{\beta\gamma}}{\partial x^{\rho}}\right)\, .
\end{equation}
The Riemann tensor is defined as
\begin{equation}
R^{\alpha}_{~\beta\mu\nu}=
\Gamma^{\alpha}_{\beta\nu,\mu}-\Gamma^{\alpha}_{\beta\mu,\nu}+
\Gamma^{\alpha}_{\lambda\mu}\Gamma^{\lambda}_{\beta\nu}-
\Gamma^{\alpha}_{\lambda\nu}\Gamma^{\lambda}_{\beta\mu} \,.
\end{equation}

The Ricci tensor is a contraction of the Riemann tensor
\begin{equation}
R_{\mu\nu}=R^{\alpha}_{~\mu\alpha\nu} \,,
\end{equation}
and in terms of the connection coefficient it is given by
\begin{equation}
R_{\mu\nu}\,=\, \partial_\alpha\,\Gamma^\alpha_{\mu\nu} \,-\,
\partial_{\mu}\,\Gamma^\alpha_{\nu\alpha} \,+\,
\Gamma^\alpha_{\sigma\alpha}\,\Gamma^\sigma_{\mu\nu} \,-\,
\Gamma^\alpha_{\sigma\nu} \,\Gamma^\sigma_{\mu\alpha}\,.
\end{equation}
The Ricci scalar is given by contracting the Ricci tensor
\begin{equation}
R=R^{\mu}_{~\mu} \,.
\end{equation}
The Einstein tensor is defined as
\begin{equation}
G_{\mu\nu}=R_{\mu\nu}-\frac{1}{2}g_{\mu\nu}R \,.
\end{equation}

\subsection{The connection coefficients}
\noindent 
For the connection coefficients we find
\begin{eqnarray}
\Gamma^0_{00}&=& {\mathcal H}+\Phi'\, ,\nonumber\\
\Gamma^0_{0i}&=& \frac{\partial\Phi}{\partial x^i}+
{\mathcal H}\omega_i\, ,\nonumber\\
\Gamma^i_{00}&=& \omega^{i'}+{\mathcal H}\omega^i+e^{2\Psi+2\Phi}
\frac{\partial\Phi}{\partial x_i}
\, ,\nonumber\\
\Gamma^0_{ij}&=& -\frac{1}{2}\left(\frac{\partial \omega_j}{\partial x^i}+
\frac{\partial \omega_i}{\partial x^j}\right)+e^{-2\Psi-2\Phi}
\left({\mathcal H}-\Psi'\right)\delta_{ij}+\frac{1}{2}\chi_{ij}'+
{\mathcal H}\chi_{ij}
\, ,\nonumber\\
\Gamma^i_{0j}&=&\left({\mathcal H}-\Psi'\right)\delta_{ij}+
\frac{1}{2}\chi_{ij}'+\frac{1}{2}\left(\frac{\partial \omega_i}{\partial x^j}-
\frac{\partial \omega_j}{\partial x^i}\right)\, ,\nonumber\\
\Gamma^i_{jk}&=&-{\cal H}\omega^i\delta_{jk}-
\frac{\partial\Psi}{\partial x^k}\delta^i_{~j}-
\frac{\partial\Psi}{\partial x^j}\delta^i_{~k}+
\frac{\partial\Psi}{\partial x_i}\delta_{jk}+ \frac{1}{2} \left(\frac{\partial\chi^i_{~j}}{\partial x^k}+
\frac{\partial\chi^i_{~k}}{\partial x^j}+\frac{\partial\chi_{jk}}{\partial x_i}
\right)\, .
\end{eqnarray}
\subsection{Einstein equations}
\noindent 
The Einstein equations are written as $G_{\mu\nu}=\kappa^2 T_{\mu\nu}$, so that $\kappa^2=8\pi G_{\rm N}$, 
where $G_{\rm N}$ is the usual Newtonian gravitational constant. They read
\begin{eqnarray}
\label{00}
G^0_{~0}&=&-\frac{e^{-2\Phi}}{a^2}\left[3{\mathcal H}^2-6{\mathcal H}\Psi'
+3(\Psi')^2-e^{2\Phi+2\Psi}\left(\partial_i\Psi\partial^i\Psi-2\nabla^2
\Psi\right)\right]= \kappa^2 T^0_{~0}\, ,\\
\label{i0}
G^i_{~0}&=&2\frac{e^{2\Psi}}{a^2}\left[\partial^i\Psi'+\left({\mathcal H}-
\Psi'\right)\partial^i\Phi\right]-\frac{1}{2a^2}\nabla^2\omega^i
+\left(4{\mathcal H}^2-2\frac{a''}{a}\right)\frac{\omega^i}{a^2}=\kappa^2 T^i_{~0}\, ,\\
\label{ij}
G^i_{~j}&=&\frac{1}{a^2}\left[e^{-2\Phi}\left({\mathcal H}^2-2\frac{a''}{a}-
2\Psi'\Phi'-3(\Psi')^2+2{\mathcal H}\left(\Phi'+2\Psi'\right)
+2\Psi''\right)\right.\nonumber\\
&+&\left. e^{2\Psi}\left(\partial_k\Phi\partial^k\Phi+\nabla^2\Phi-\nabla^2\Psi
\right)\right]\delta^i_j+\frac{e^{2\Psi}}{a^2}
\left(-\partial^i\Phi\partial_j\Phi-
\partial^i\partial_j\Phi+\partial^i\partial_j\Psi-\partial^i\Phi\partial_j\Psi
+\partial^i\Psi\partial_j\Psi-\partial^i\Psi\partial_j\Phi\right)\nonumber\\
&-&\frac{{\mathcal H}}{a^2}\left(\partial^i\omega_j+\partial_j\omega^i\right)
-\frac{1}{2a^2}\left(\partial^{i}\omega_j'+\partial_j\omega^{i'}\right)
+\frac{1}{a^2}\left(
{\mathcal H}\chi^{i'}_j+\frac{1}{2}\chi_j^{i''}-\frac{1}{2}\nabla^2
\chi^i_j\right)=\kappa^2 T^i_{~j}\, .
\end{eqnarray}
Here $T^\mu_{~\nu}$ is the energy momentum tensor accounting for different components, photons, baryons, dark matter. 
We will give the expressions later for each component in terms
 of the distribution functions.  

\section{The collisionless Boltzmann equation for photons}
\noindent 
We are now interested in the anisotropies in the cosmic
distribution of photons and inhomogeneities in the matter. Photons
are affected by gravity and by Compton scattering
with free electrons. The latter are tightly coupled to protons. Both
are, of course, affected by gravity. The metric which
determines the gravitational forces is influenced by all these components
plus CDM (and neutrinos). Our plan is to write down Boltzmann equations for 
the phase-space distributions of each species in the Universe.

The phase-space distribution of particles $g(x^i,P^\mu,\eta)$ 
is a function of spatial coordinates $x^i$, conformal time $\eta$, and 
momentum of the particle 
$P^\mu=dx^\mu/d\lambda$ where $\lambda$ parametrizes the particle path. 
Through the constraint $P^2 \equiv g_{\mu\nu} P^\mu P^\nu =- m^2$, where $m$ is 
the mass of the particle one can eliminate $P^0$ and $g(x^i,P^j,\eta)$ 
gives the number of particles in a differential volume 
$dx^1 dx^2 dx^3 dP^1 dP^2 dP^3$ in phase space. In the following we will indicate the 
distribution function for photons with $f$. 

The photons' distribution evolves according to the Boltzmann equation 
\begin{equation}
\label{Boltzgeneric}
\frac{df}{d\eta}=a\, C[f]\, ,
\end{equation}  
where the collision term is due to scattering of photons off free electrons. In the following we will derive the left-hand side of 
Eq.~(\ref{Boltzgeneric}) while in the next section we will compute the collision term.

For photons we can impose $P^2 \equiv g_{\mu\nu} P^\mu P^\nu =0$ and using 
the metric~(\ref{metric}) in the conformal time $\eta$ we find
\begin{equation}
\label{P=0}
P^2=a^2\left[ - e^{2\Phi} (P^0)^2+\frac{p^2}{a^2}+2 \omega_i P^0 P^i\right]=0\, ,
\end{equation} 
where we define
\begin{equation}
\label{defp}
p^2= g_{ij}  P^iP^j\, .
\end{equation}
From the constraint~(\ref{P=0})
\begin{equation}
\label{P0}
P^0=e^{-\Phi}\left( \frac{p^2}{a^2}+2 \omega_i P^0P^i\right )^{1/2}\, .
\end{equation}
Notice that we immediately recover the usual zero and first-order relations
$P^0=p/a$ and $P^0=p(1-\Phi^{(1)})/a$. 

The components $P^i$ are proportional to $p n^i$, where $n^i$ is a unit 
vector with $n^in_i = \delta_{ij} n^in^j=1$.  
We can write $P^i=C n^i$, where $C$ is determined by
\begin{equation}
g_{ij} P^iP^j=C^2 \, a^2 (e^{-2\Psi}+\chi_{ij}n^in^j)=p^2\, ,
\end{equation}
so that 
\begin{equation}
\label{Pi}
P^i=\frac{p}{a} n^i\left(e^{-2\Psi}+\chi_{km}n^kn^m \right)^{-1/2} = 
\frac{p}{a} n^i e^{\Psi} \left( 1-\frac{1}{2} \chi_{km}n^kn^m \right)\, , 
\end{equation}
where the last equality holds up to second order in the perturbations. Again 
we recover the zero and first-order relations $P^i=p n^i/a$ and 
$P^i=p n^i (1+\Psi^{(1)})/a$ respectively. Thus up to second order we can write
\begin{equation}
\label{P0bis}
P^0=e^{-\Phi} \frac{p}{a} \left( 1+\omega_i\, n^i \right)\, .
\end{equation} 
Eq.~(\ref{Pi}) and~(\ref{P0bis}) allow us to replace $P^0$ and $P^i$ in terms 
of the variables $p$ and $n^i$. Therefore, as 
it is standard in the literature, from now on we will consider the phase-space 
distribution $f$ as a function of the momentum ${\bf p}=p n^i$ with magnitude 
$p$ and angular direction $n^i$, $f\equiv f(x^i, p, n^i,\eta)$. 

 Thus in terms 
of these variables  the total time derivative of the distribution function 
reads
\begin{equation}
\label{Df}
\frac{d f}{d \eta}=\frac{\partial f}{\partial \eta}+
\frac{\partial f}{\partial x^i} \frac{d x^i}{d \eta}+
\frac{\partial f}{\partial p} \frac{d p}{d \eta}+
\frac{\partial f}{\partial n^i} \frac{d n^i} {d \eta}\, .
\end{equation}
In the following we will compute $d x^i/d \eta$, $d p/d \eta$ and 
$d n^i/d \eta$: 
\\
\\
a) $d x^i/d \eta$: 
\\
\\
From 
\begin{equation}
P^i=\frac{d x^i}{d \lambda}=\frac{d x^i}{d \eta} \frac{d\eta}{d\lambda}=\frac{d x^i}{d \eta} P^0
\end{equation}
and from Eq.~(\ref{Pi}) and~(\ref{P0bis}) 
\begin{equation}
\label{dxi}
\frac{d x^i}{d \eta}= n^i e^{\Phi+\Psi}
\left( 1- \omega_{j}\, n^j - \frac{1}{2} 
\chi_{km} n^k n^m \right)\, .
\end{equation}   
\\
b) $d p/d \eta$:
\\
\\
For $dp/d \eta$ we make use of the time component of the geodesic equation 
$d P^0/d \lambda= - \Gamma^0_{\alpha \beta} P^{\alpha} P^{\beta}$ where we can replace the derivative 
$d/d\lambda$ with $(d\eta/d\lambda) 
\, d/d\eta=P^0 \, d/d\eta$ so that  
\begin{equation}
\label{0geod}
\frac{d P^0}{ d \eta}= - \Gamma^0_{\alpha \beta} 
\frac{P^{\alpha} P^{\beta}}{P^0}\, ,
\end{equation}
Using the metric~(\ref{metric}) we find 
\begin{eqnarray}
\label{GPP}
2 \Gamma^0_{\alpha \beta} P^{\alpha} P^{\beta}&=& g^{0\nu} 
\left[ 2 \frac{\partial g_{\nu \alpha}}{\partial x^\beta} 
-\frac{\partial_{\alpha \beta}}{\partial x^\nu}
\right]P^\alpha P^\beta \nonumber \\
&=&2 ({\mathcal H}+\Phi')\left( P^0 \right)^2+4 \Phi_{,i} P^0P^i 
+4 {\mathcal H} \omega_i P^0 P^i+
2 e^{-2\Phi}\left[ ({\cal H}-\Psi') e^{-2 \Psi} \delta_{ij}  
-\omega_{i,j} + \frac{1}{2} \chi'_{ij}+{\cal H}\chi_{ij} \right] P^iP^j \, .
\nonumber \\
\end{eqnarray}
On the other hand the expression~(\ref{P0bis}) of $P^0$ in terms of $p$ and $n^i$ gives
\begin{equation}
\frac{dP^0}{d \eta}=-\frac{p}{a} \frac{d \Phi}{d \eta} e^{-\Phi} 
\left( 1+ \omega_i n^i
\right)+e^{-\Phi} \left( 1+\omega_i \, n^i \right) \frac{d(p/a)}{d\eta}+
\frac{p}{a} e^{-\Phi} \frac{d(\omega_i \, n^i)}{d\eta}\, .
\end{equation} 
Thus Eq.~(\ref{0geod}) allows us express $dp/d\eta$ as
\begin{eqnarray}
\label{dp1}
\frac{1}{p} \frac{dp}{d\eta}=- {\cal H} +\Psi'-\Phi_{,i} \,n^i e^{\Phi+\Psi} 
- \omega'_{i}\,n^i-\frac{1}{2} \chi'_{ij}n^in^j\, ,
\end{eqnarray} 
where in Eq.~(\ref{GPP}) we have replaced $P^0$ and $P^i$ by 
Eqs.~(\ref{P0bis}) and 
(\ref{Pi}). Notice that in order to obtain Eq.~(\ref{dp1}) 
we have used the following 
expressions for the total time derivatives of the metric perturbations
\begin{eqnarray}
\label{dPhi}
\frac{d \Phi}{d\eta}=\frac{\partial \Phi}{\partial \eta}+
\frac{\partial \Phi}{\partial x^i} \frac{d x^i}{d \eta}= 
\frac{\partial \Phi}{\partial \eta}+\frac{\partial \Phi}{\partial x^i}n^i 
e^{\Phi+\Psi}\left( 1-\omega_j\, n^j -\frac{1}{2} \chi_{km}n^k n^m \right)\, ,
\end{eqnarray}
and 
\begin{eqnarray}
\frac{d (\omega_{i} n^i)}{d \eta}=n^i\left( 
\frac{\partial \omega_i}{\partial \eta}+
\frac{\partial \omega_i}{\partial x^j} \frac{dx^j}{d\eta}\right)=
\frac{\partial \omega_i}{\partial \eta}n^i+
\frac{\partial \omega_i}{\partial x^j} n^i n^j \, ,
\end{eqnarray}
where we have taken into account that $\omega_i$ is already a second-order 
perturbation 
so that we can neglect $dn^i/d\eta$ which is at least a first order quantity, 
and we can 
take the zero-order expression in Eq.~(\ref{dxi}), $dx^i/d\eta=n^i$.     
In fact there is also an alternative expression for 
$dp/d\eta$ which turns out to be useful later and which can be obtained by 
applying once more Eq.~(\ref{dPhi}) 
\begin{eqnarray}
\label{dp2}
\frac{1}{p}\frac{dp}{d\eta}=-{\cal H} -\frac{d\Phi}{d\eta}
+\Phi'+\Psi'-\omega'_{i}\,n^i
-\frac{1}{2}\chi'_{ij} n^in^j\, .
\end{eqnarray}
\\ 
\\
c) $d n^i/d \eta$: 
\\
\\
We can proceed in a similar way to compute $dn^i/d\eta$. Notice that since in Eq.~(\ref{Df}) it multiplies 
$\partial f/\partial n^i$ which is first order, we need only the first order perturbation of $dn^i/d\eta$.  
We use the spatial components of the geodesic equations $dP^i/d\lambda=- 
\Gamma^i_{\alpha \beta}P^{\alpha}P^{\beta}$ written as
\begin{eqnarray}
\label{geoi}
\frac{dP^i}{d\eta}=-\Gamma^i_{\alpha \beta}\frac{P^{\alpha}P^{\beta}}{P^0}\, .
\end{eqnarray}
For the right-hand side we find up to second-order
\begin{eqnarray}
\label{RHSgeoi}
2\Gamma^i_{\alpha \beta} P^{\alpha}P^{\beta}&=&g^{i\nu} \left[ 
\frac{\partial g_{\alpha\nu}}{\partial x^\beta} +\frac{\partial g_{\beta \nu}}{\partial x^{\alpha}}
-\frac{\partial g_{\alpha \beta}}{\partial x^\nu}\right] P^\alpha P^\beta\nonumber \\
&=& 4 ({\cal H}-\Psi')P^i P^0 +2\left( \chi^{i\prime}_{~k}+\omega^i_{,k}-\omega_k^{~,i} \right) P^0P^k +
\left(2 \frac{\partial \Phi}{\partial x^i} e^{2\Phi+2\Psi} +2 \omega^{i\prime} +2 {\cal H} \omega^i \right)
\left( P^0 \right)^2 \nonumber - 4 \frac{\partial \Psi}{\partial x^k} P^i P^k \nonumber \\
&+&2 \frac{\partial \Psi}{\partial x^i}\delta_{km} P^kP^m
- \left[ 2{\cal H} \omega^i \delta_{jk}-\left(\frac{\partial\chi^i_{~j}}{\partial x^k}+
\frac{\partial\chi^i_{~k}}{\partial x^j}+\frac{\partial\chi_{jk}}{\partial x_i }\right) \right]P^jP^k
\, , 
\end{eqnarray}
while the expression~(\ref{Pi}) of $P^i$ in terms of our variables $p$ 
and $n^i$ in the left-hand side of Eq.~(\ref{geoi}) brings
\begin{eqnarray}
\frac{dP^i}{d\eta}=\frac{p}{a} e^{\Psi}\left[ \frac{d\Psi}{d\eta} n^i+
\frac{a}{p} \frac{d(p/a)}{d\eta} n^i+\frac{dn^i}{d\eta} \right]
\left(1-\frac{1}{2} \chi_{km} n^kn^m \right) 
-\frac{p}{a}n^i e^{\Psi} \frac{1}{2} \frac{d \left(\chi_{km}n^kn^m\right)}{d\eta} \, . 
\end{eqnarray}
Thus, using the expression~(\ref{Pi}) for $P^i$ and~(\ref{P0}) for $P^0$ in 
Eq.~(\ref{RHSgeoi}), together with the previous result~(\ref{dp1}), 
the geodesic equation~(\ref{geoi}) gives the following 
expression $dn^i/d\eta$ (valid up to first order)
\begin{eqnarray}
\label{dni}
\frac{d n^i}{d\eta}=\left( \Phi_{,k}+
\Psi_{,k} \right) n^k n^i-\Phi^{,i}-\Psi^{,i}\, .
\end{eqnarray}

To proceed further we now expand the distribution function for 
photons around the zero-order value $f^{(0)}$ which is that of a Bose-Einstein 
distribution
\begin{equation}
\label{BEd}
f^{(0)}(p,\eta)=2\,\, \frac{1}{\exp\left\{\frac{p}{T(\eta)}\right\}-1}\, ,
\end{equation}
where $T(\eta)$ is the average (zero-order) temperature and the factor $2$ 
comes from the spin degrees of photons. The perturbed 
distribution of photons will depend also on $x^i$ and on the propagation 
direction $n^i$ so as to account for inhomogeneities and anisotropies
\begin{equation}
\label{expf}
f(x^i,p,n^i,\eta)=f^{(0)}(p,\eta)+f^{(1)}(x^i,p,n^i,\eta)+\frac{1}{2} 
f^{(2)}(x^i,p,n^i,\eta)\, ,
\end{equation} 
where we split the perturbation of the distribution function into a 
first and a second-order part. The Boltzmann equation 
up to second-order can be written in a straightforward way by 
recalling that the total time derivative of a given $i$-th perturbation, as 
{\it e.g.} $df^{(i)}/d\eta$ is {\it at least} a quantity of the $i$-th order. 
Thus it is easy to realize, looking at Eq.~(\ref{Df}), that 
 the left-hand side of Boltzmann equation can be written 
{\it up to second-order} as 
\begin{eqnarray}
\label{LHSBoltz}
\frac{df}{d\eta}&=&
\frac{d f^{(1)}}{d\eta}+\frac{1}{2} \frac{df^{(2)}}{d\eta}
-p\frac{\partial f^{(0)}}{\partial p}\frac{d}{d\eta}\left(\Phi^{(1)}+\frac{1}{2} 
\Phi^{(2)}\right) +
p \frac{\partial f^{(0)}}{\partial p} \frac{\partial }{\partial \eta}
\left( \Phi^{(1)}+\Psi^{(1)}+\frac{1}{2} \Phi^{(2)}+\frac{1}{2} \Psi^{(2)} \right) \nonumber \\
&-& p \frac{\partial f^{(0)}}{\partial p} \frac{\partial \omega_i}{\partial \eta}n^i
- \frac{1}{2} p \frac{\partial f^{(0)}}{\partial p} \frac{\partial \chi_{ij}} 
{\partial \eta} n^in^j\, ,
\end{eqnarray}
where for simplicity in Eq.~(\ref{LHSBoltz}) we have already used the 
background Boltzmann equation $(df/d\eta)|^{(0)}=0$. 
In Eq.~(\ref{LHSBoltz}) there are all the terms which will give rise to the integrated Sachs-Wolfe 
effects (corresponding to the terms which explicitly 
depend on the gravitational perturbations), 
while other effects, such as the gravitational lensing, are still 
hidden in the (second-order part) of the first term. In fact in order to 
obtain Eq.~(\ref{LHSBoltz}) we just need for the time being  
to know the expression for $dp/d\eta$, Eq.~(\ref{dp2}). 
\section{Collision term}
\noindent 
\subsection{The Collision Integral}
\label{CI}
\noindent 
In this section we focus on the collision term due to Compton scattering 
\begin{equation}
e({\bf q}) \gamma({\bf p}) \longleftrightarrow e({\bf q}') 
\gamma({\bf p}')\, ,
\end{equation}
where we have indicated the momentum of the photons  and electrons involved in 
the collisions. The collision term will be important for small scale anisotropies and 
spectral distortions. The important point to compute the collision term is that for 
the epoch of interest very little energy is transferred. Therefore one can 
proceed by expanding the right hand side of Eq.~(\ref{Boltzgeneric}) 
both in the small perturbation, Eq.~(\ref{expf}), and in the small energy 
transfer. Part of the computation up to second-order has been done in 
Refs.~\cite{HSS,DJ,Huthesis} (see also \cite{roy}). In particular 
Refs.~\cite{HSS,DJ} are focused on the effects 
of reionization on the CMB anisotropies thus keeping in the collision term those 
contributions relevant for the small scale effects due to reionization and neglecting 
the effects of the metric perturbations on the left-hand side of Eq.~(\ref{Boltzgeneric}). 
We will mainly follow the formalism of Ref.~\cite{DJ} and we will keep all the terms arising 
from the expansion of the collision term up to second-order.   

The collision term is given by
\begin{eqnarray}
\label{collisionterm}
C({\bf p})&=&\frac{1}{E({\bf p})} \int \frac{d{\bf q}}{(2 \pi)^3 2E({\bf q})} 
\frac{d{\bf q}'}{(2 \pi)^3 2E({\bf q}')} \frac{d{\bf p}'}{(2 \pi)^3 2E({\bf p}')}
(2\pi)^4 \delta^4(q+p-q'-p') \left| M \right|^2  \nonumber \\
& \times & \{ g({\bf q}')f({\bf p}')[ 1+f({\bf p})]-
g({\bf q})f({\bf p})[ 1+f({\bf p}')]\} 
\end{eqnarray}
where $E({\bf q})=(q^2+m_e^2)^{1/2}$, $M$ is the amplitude of the scattering process,
$\delta^4(q+p-q'-p')=\delta^3({\bf q}+{\bf p}-{\bf q}'-{\bf p}') 
\delta(E({\bf q})+p-E({\bf q}') -p')$ 
ensures the energy-momentum conservation and $g$ is 
the distribution function for electrons. The Pauli suppression factors $(1-g)$
 have been 
dropped since for the epoch of interest the density of electrons $n_e$ is low. 
The electrons are kept in thermal equilibrium by the Coulomb interactions with protons and 
they are non-relativistic, thus we can take a Boltzmann distribution around some bulk velocity ${\bf v}$
\begin{eqnarray}
\label{gel}
g({\bf q})=n_e \left( \frac{2 \pi}{m_e T_e}\right)^{3/2} 
\exp\left\{-\frac{({\bf q}-m_e{\bf v})^2}{2m_e T_e} \right\} 
\end{eqnarray}      
By using the three dimensional delta function the energy transfer is given by 
$E({\bf q})-E({\bf q}+{\bf p}-{\bf p}')$ and it turns out to be small compared to the 
typical thermal energies
\begin{equation}
\label{par}
E({\bf q})-E({\bf q}+{\bf p}-{\bf p}')\simeq \frac{({\bf p}-{\bf p}') \cdot {\bf q}}{m_e}
={\cal O}(Tq/m_e)\, . 
\end{equation}  
In Eq.~(\ref{par}) we have used $E({\bf q})=m_e+q^2/2m_e$ and the fact that since the scattering 
is almost elastic ($p\simeq p'$) $({\bf p}-{\bf p'})$ is of order $p\sim T$, with $q$ much bigger 
than $({\bf p}-{\bf p'})$. In general, 
the electron momentum has two contributions, 
the bulk velocity ($q=m_e v$)  and the thermal 
motion ($q \sim (m_e T)^{1/2}$) and thus the parameter expansion $q/m_e$ includes 
the small bulk velocity ${\bf v}$ and the the ratio $(T/m_e)^{1/2}$ which is small because 
the electrons are non-relativistic. 

The expansion of all the quantities entering the collision term 
in the energy transfer parameter and the integration over the momenta ${\bf q}$ and 
${\bf q}'$ is described in details in Ref.~\cite{DJ}. 
Here we just note that it is easy to realize that we just need the scattering amplitude 
up to first order since at zero order 
$g({\bf q}')=g({\bf q}+{\bf p}-{\bf p}')=g({\bf q})$ 
and $\delta(E({\bf q})+p-E({\bf q}')-p')=\delta(p-p')$ so that all the zero-order quantities 
multiplying $\left| M \right|^2$ vanish. To first order 
\begin{equation}
\left| M \right|^2=6\pi\sigma_T m_e^2[(1+\cos^2\theta)-2\cos\theta(1-\cos\theta){\bf q}\cdot 
({\bf \hat{p}}+{\bf \hat{p}}')/m_e]\, ,
\end{equation}
where $\cos\theta={\bf n} \cdot {\bf n'}$ is the scattering angle and $\sigma_T$ the 
Thompson cross section. The resulting collision term up to second order is given by~\cite{DJ} 
\begin{eqnarray}
\label{Integralcolli}
C(\bp) =  
 {3n_e\sigma_T\over 4p} \int dp' p' {d\Omega' \over 4 \pi}
\bigg[ c^{(1)}(\bp,\bpp) + c^{(2)}_\Delta(\bp,\bpp)+c^{(2)}_v(\bp,\bpp)
+ c^{(2)}_{\Delta v}(\bp,\bpp) + c^{(2)}_{vv}(\bp,\bpp)+c^{(2)}_K(\bp,\bpp)   \bigg]\, ,
\end{eqnarray}
where we arrange the different contributions following Ref.~\cite{DJ}.
The first order term reads 
\begin{eqnarray}
c^{(1)}(\bp,\bpp) = (1+\cos^2\theta)\Bigg[
\delta(p-p') (\fps1-\f1)
+(\fpz-\fz) (\bp-\bpp)\cdot\bv {\partial \delta(p-p')
			\over \partial p'} \Bigg]\, ,
\end{eqnarray}
while  the second-order terms  have been separated into four parts. 
There is the so-called anisotropy suppression term 
\begin{eqnarray}
c^{(2)}_\Delta(\bp,\bpp) =\frac{1}{2} \left(1+\cos^2\theta\right)
		 \delta(p-p')(\fps2 - \f2)\, ;
\end{eqnarray}
a term which depends on the second-order velocity 
perturbation defined by the expansion 
of the bulk flow as ${\bf v}={\bf v}^{(1)}+{\bf v}^{(2)}/2$
\begin{equation}
c^{(2)}_v(\bp,\bpp)=\frac{1}{2}(1+\cos^2 \theta)  
(\fpz-\fz) (\bp-\bpp)\cdot\bv^{(2)}\, {\partial \delta(p-p')
			\over \partial p'}\, ;
\end{equation}
a set of terms coupling the photon perturbation to the velocity
\begin{eqnarray}
 c^{(2)}_{\Delta v}(\bp,\bpp)=
\left(\fps1 - \f1\right)
\Bigg[  \left(1+\cos^2\theta\right)(\bp-\bpp)\cdot\bv {\partial \delta(p-p')
			\over \partial p'}-2\cos\theta(1-\cos\theta)
	\delta(p-p')
		({\bf n}+ {\bf n}')\cdot \bv
\Bigg]\, , \nonumber \\
\end{eqnarray}
and a set of source terms quadratic in the velocity
\begin{eqnarray}
c^{(2)}_{vv}(\bp,\bpp) & = &
	\left(\fpz-\fz\right)\ (\bp-\bpp)\cdot\bv
\Bigg[
	\left(1+\cos^2\theta\right) {(\bp-\bpp)\cdot\bv\over2}
		 {\partial^2 \delta(p-p')
			\over \partial p'^2} \nonumber \\
&-& 2\cos\theta(1-\cos\theta)
({\bf n}+ {\bf n}')\cdot \bv{\partial \delta(p-p')
			\over \partial p'} \Bigg]\,. \nonumber \\
\end{eqnarray}
The last contribution are the Kompaneets terms 
describing spectral distortions to the CMB
\begin{eqnarray}
c^{(2)}_K(\bp,\bpp)& = & \left(1+\cos^2\theta\right) {(\bp-\bpp)^2\over2\me}
	\Bigg[  \left( \fpz-\fz\right) \te {\partial^2 \delta(p-p')
			\over \partial p'^2} \nonumber \\ 
&-& \left(\fpz+\fz+2\fpz\fz\right) {\partial \delta(p-p')
			\over \partial p'} \Bigg] \nonumber \\
&+& {2(p-p')\cos\theta(1-\cos^2\theta)\over\me}
	\Bigg[ \delta(p-p') \fpz(1+\fz)
\left(\fpz-\fz\right){\partial \delta(p-p')
			\over \partial p'} \Bigg]\, .
\end{eqnarray}
Let us make a couple of comments about the various contributions to the collision term. First, 
notice the term $c^{(2)}_v(\bp,\bpp)$ due to second-order 
perturbations in the velocity of electrons which is absent in Ref.~\cite{DJ}. 
In standard cosmological scenarios (like inflation) vector perturbations are not generated at 
linear order, so that linear velocities are irrotational $v^{(1)i}=\partial^i v^{(1)}$. However at 
second-order vector perturbations are generated after horizon crossing as non-linear combinations 
of primordial scalar modes. Thus we must take into account
also a transverse (divergence free) component, $v^{(2)i}=\partial^i v^{(2)}+ v^{(2)i}_T$ with 
$\partial_i v^{(2)i}_{T}=0$. As we will see such vector perturbations will break azimuthal symmetry 
of the collision term with respect to a given mode ${\bf k}$, which instead  usually holds at linear order. Secondly, notice that the 
number density of electrons appearing in Eq.~(\ref{Integralcolli}) must be expanded as 
$n_e = \bar{n}_e(1+\delta_e)$ and then 
\begin{equation}
\label{deltac1}
\delta^{(1)}_e \,c^{(1)}(\bp,\bpp)
\end{equation}
gives rise to second-order contributions in addition to the list above, where we split $\delta_e=\delta^{(1)}_e+\delta^{(2)}_e/2$ into a 
first- and second-order part. In particular the 
combination with the term proportional to ${\bf v}$ in $c^{(1)}(\bp,\bpp)$ gives rise to the so 
called Vishniac effect as discussed in Ref.~\cite{DJ}.  
\subsection{Computation of different contributions to the collision term}
In the integral~(\ref{Integralcolli}) over the momentum ${\bf p}'$ the first-order term gives the 
usual collision term
\begin{equation}
\label{C1}
C^{(1)}({\bf p})=n_e \sigma_T \left[ f^{(1)}_0(p)+\frac{1}{2}f^{(1)}_2 P_2({\bf \hat v} \cdot 
{\bf n})-f^{(1)}-p\frac{\partial f^{(0)}}{\partial p} {\bf v} \cdot {\bf n} 
\right]\, ,
\end{equation}
where one uses the decomposition in Legendre polynomials
\begin{equation} 
\label{dec1}
f^{(1)}({\bf x},p,{\bf n})=\sum_\ell (2\ell +1) f^{(1)}_\ell(p) P_\ell(\cos \vartheta)\, ,
\end{equation}
where $\vartheta$ is the polar angle of ${\bf n}$, $\cos \vartheta ={\bf n} \cdot {\bf \hat{v}}$.

In the following we compute the second-order collision term separately for the different 
contributions, using the notation $C(\bp)=C^{(1)}(\bp)+C^{(2)}(\bp)/2$. 
We have not reported the details of the calculation of the first-order term because 
for its second-order analog, $c^{(2)}_{\Delta}(\bp,\bpp)+c^{(2)}_v(\bp,\bpp)$, 
the procedure is the same. The important 
difference is that the second-order velocity term includes a vector part, and this leads to 
a generic angular decomposition of the distribution function (for simplicity drop the time 
dependence) 
\begin{equation}
\label{fangdeco}
f^{(i)}({\bf x},p,{\bf n})=\sum_{\ell} \sum_{m=-\ell}^{\ell} f^{(i)}_{\ell m}({\bf x},p)  
(-i)^{\ell}\ \sqrt{\frac{4\pi}{2\ell+1}} Y_{\ell m}({\bf n})\, ,
\end{equation} 
such that 
\begin{equation}
\label{angular}
f^{(i)}_{\ell m}=(-i)^{- \ell}\sqrt{\frac{2\ell+1}{4\pi}} \int d\Omega  f^{(i)} 
Y^{*}_{\ell m}({\bf n}) \, .
\end{equation}
Such a decomposition holds also in Fourier space \cite{Complete}.
The notation at this stage is a bit confusing, so let us restate it:
superscripts 
denote the order of the perturbation; the subscripts refer to the moments
of the distribution.  Indeed at first order one can drop the dependence on $m$ 
setting $m=0$ using the fact that the distribution function does not depend on the azimuthal angle 
$\phi$. In this case the relation with $f^{(1)}_l$ is 
\begin{equation}
\label{rel}
f^{(1)}_{\ell m}=(-i)^{-\ell} (2\ell +1)  \delta_{m0} \, f^{(1)}_{\ell}\, .
\end{equation} 
\\
\\
a) \quad $c^{(2)}_\Delta(\bp, \bpp)$
\\
\\
The integral over $\bpp$ yields
\begin{eqnarray}
C^{(2)}_\Delta(\bp)=\frac{3n_e \sigma_T}{4p}\int dp' p' \frac{d \Omega'}{4 \pi} c^{(2)}_\Delta(\bp, \bpp) &=&
\frac{3n_e \sigma_T}{4p}\int dp' p'\delta(p-p') \int \frac{d \Omega'}{4 \pi} 
[1+({\bf n} \cdot {\bf n}')^2] [f^{(2)}(\bpp)-f^{(2)}(\bp)]\, . \nonumber \\
\end{eqnarray}
To do the angular integral we write the angular dependence on the scattering angle 
$\cos \theta= {\bf n} \cdot {\bf n}'$ in terms of the Legendre polynomials  
\begin{equation}
\label{DL}
[1+({\bf n} \cdot {\bf n}')^2]=\frac{4}{3}\left[1+\frac{1}{2} P_2({\bf n} \cdot {\bf n}') \right]=
\left[1+\frac{1}{2}\sum_{m=-2}^{2} Y_{2m}({\bf n})  Y^{*}_{2m}({\bf n}') \frac{4 \pi}{2\ell +1}
\right] \, ,
\end{equation}
where in the last step we used the addition theorem of spherical harmonics
\begin{equation}
P_\ell=\frac{4 \pi}{2\ell +1} \sum_{m=-2}^{2} Y_{\ell m}({\bf n})  Y^{*}_{\ell m}({\bf n}')\, . 
\end{equation}
Using the decomposition~(\ref{angular}) and the orthonormality of the spherical harmonics we find
\begin{equation}
C^{(2)}_\Delta(\bp)=n_e \sigma_T \left[ 
f^{(2)}_{0 0}(p)-f^{(2)}(\bp)-\frac{1}{2} \sum_{m=-2}^{2} 
\frac{\sqrt{4 \pi}}{5^{3/2}}\, f^{(2)}_{2m}(p) \, Y_{2m}({\bf n}) \right] \, .  
\end{equation} 
It is easy to recover the result for the corresponding first-order contribution in Eq.~(\ref{C1})
by using Eq.~(\ref{rel}).
\\
\\
b) \quad $c^{(2)}_v(\bp,\bpp)$
\\
\\
Let us fix for simplicity our coordinates such that the polar angle of ${\bf n}'$ is defined by $\mu'=
{\bf \hat{v}}^{(2)} \cdot {\bf n}'$ with $\phi'$ the corresponding azimuthal angle. The contribution 
of  $c^{(2)}_v(\bp,\bpp)$ to the collision term is then 
\begin{equation}
C^{(2)}_v(\bp)= \frac{3 n_e \sigma_T}{4 p} v^{(2)} \int dp' p' 
[f^{(0)}(p')-f^{(0)}(p)]\frac{\partial \delta(p-p')}{\partial p'} 
\int_{-1}^{1} \frac{d \mu'}{2} (p \mu-p'\mu') \int_0^{2\pi} \frac{d \phi'}{2\pi} 
[1+({\bf p}\cdot {\bf p'})^2]\, .
\end{equation} 
We can use Eq.~(\ref{DL}) which in our coordinate system reads
\begin{equation}
\label{DL2}
[1+({\bf n} \cdot {\bf n}')^2]=\frac{4}{3}\left[1+\frac{1}{2} P_2({\bf n} \cdot {\bf n}') \right]=
\frac{4}{3}\left[1+\frac{1}{2}  
\sum_{m=-2}^m \frac{(2-m)!}{(2+m)!} P_2^m({\bf n}\cdot{\bf \hat{ v}}^{(2)}) P_2^m({\bf n}' \cdot{\bf\hat{ v}}^{(2)})
e^{im(\phi'-\phi)} \right] \, ,
\end{equation}
so that 
\begin{equation}
\label{intphi}
\int \frac{d \phi'}{2 \pi} P_2({\bf n} \cdot{\bf n}') = 
P_2({\bf n} \cdot{\bf {\hat v}}^{(2)})  P_2({\bf n}' \cdot{\bf {\hat v}}^{(2)}) =P_2(\mu)P_2(\mu')\, .
\end{equation}
By using the orthonormality of the Legendre polynomials and integrating by parts over $p'$ we find 
\begin{equation}
C^{(2)}_v(\bp)= - n_e\,  \sigma_T\,
p \frac{\partial f^{(0)}}{\partial p} {\bf v}^{(2)} \cdot {\bf n}\, . 
\end{equation}
As it is clear by the presence of the scalar product ${\bf v}^{(2)} \cdot {\bp}$ the final result is 
independent of the coordinates choosen.
\\
\\
c) \quad $c^{(2)}_{\Delta v}(\bp,\bpp)$
\\
\\
Let us consider the contribution from the first term
\begin{equation}
c^{(2)}_{\Delta v(I)}(\bp,\bpp)=\left(1+\cos^2\theta\right) \left(\fps1 - \f1\right)
(\bp-\bpp)\cdot\bv {\partial \delta(p-p')
			\over \partial p'}\, ,
\end{equation}
where the velocity has to be considered at first order. 
In the integral~(\ref{Integralcolli}) it brings
\begin{equation}
\frac{1}{2}C^{(2)}_{\Delta v (I)}= \frac{3 n_e \sigma_T v}{4 p} \int dp' p' 
\frac{\partial \delta(p-p')}{\partial p'} 
\int_{-1}^1 \frac{d \mu'}{2} [f^{(1)}(\bpp)-f^{(1)}(\bp)] (p\mu - p'\mu') 
\int_0^{2\pi} \frac{d \phi'}{2 \pi} (1+\cos^2 \theta) \, ,
\end{equation} 
The procedure to do the integral is the same as above. We use the same relations as in
Eqs.~(\ref{DL2}) and~(\ref{intphi}) where now the angles are those taken with respect to 
the first-order velocity. This eliminates the integral over $\phi'$, and integrating by parts 
over $p'$ yields
\begin{equation}
\label{interm}
\frac{1}{2} 
C^{(2)}_{\Delta v (I)}(\bp)=-\frac{3 n_e \sigma_T v}{4p} \int_{-1}^1 \frac{d \mu'}{2} 
\left[
\frac{4}{3}+\frac{2}{3} P_2(\mu) P_2(\mu') 
\right] 
\left[ 
p(\mu-2\mu')(f^{(1)}(p,\mu')-f^{(1)}(p,\mu))
+p^2(\mu-\mu') \frac{\partial f^{(1)}(p,\mu')}{\partial p} 
\right ]\, .
\end{equation} 
We now  use the decomposition~(\ref{dec1}) and the orthonormality of the Legendre polynomials to find
\begin{eqnarray}
\int \frac{d\mu'}{2} \mu' f^{(1)}(p,\mu') P_2(\mu')&=&
\sum_\ell \int \frac{d\mu'}{2} \mu ' P_2(\mu') P_l(\mu') f^{(1)}_\ell(p)=
\sum_\ell \int \frac{d\mu'}{2} \left[ \frac{2}{5} P_1(\mu') +\frac{3}{5} P_3(\mu') \right] 
P_\ell(\mu')f^{(1)}_\ell(p) \nonumber \\
&=& \frac{2}{5} f^{(1)}_1(p)+\frac{3}{5} f^{(1)}_3(p)\, ,
\end{eqnarray}
where we have used $\mu ' P_2(\mu') P_l(\mu')=\frac{2}{5} P_1(\mu') +\frac{3}{5} P_3(\mu')$, with 
$P_1(\mu')=\mu'$. Thus from Eq.~(\ref{interm}) we get 
\begin{eqnarray}
\frac{1}{2} C^{(2)}_{\Delta v (I)}(\bp)&=& n_e \sigma_T \Bigg\{ 
{\bf v} \cdot {\bf n} \left[ 
f^{(1)}(\bp) -f^{(1)}_0(p) - p\frac{\partial f^{(1)}_0(p)}{\partial p} -
\frac{1}{2} P_2({\bf \hat{v}} \cdot {\bf n}) \left( f^{(1)}_2(p)+
p\frac{\partial f^{(1)}_2(p)}{\partial p}
\right) \right] \nonumber \\
&+& v \left[ 2f^{(1)}_1(p)+ p\frac{\partial f^{(1)}_1(p)}{\partial p} + \frac{1}{5} 
P_2({\bf \hat{v}}\cdot {\bf n}) \left( 2f^{(1)}_1(p)+ p\frac{\partial f^{(1)}_1(p)}{\partial p}+
3 f^{(1)}_3(p)+\frac{3}{2} p\frac{\partial f^{(1)}_3(p)}{\partial p}
\right)
\right]
\Bigg\} \, .
\end{eqnarray}
In $c^{(2)}(\bp,\bpp)$ there is a second term
\begin{equation}
c^{(2)}_{\Delta v(II)}= 
-2\cos\theta(1-\cos\theta) 
\left( f^{(1)}(\bpp)-f^{(1)}(\bp) \right)
\delta(p-p') ({\bf n}+ {\bf n}')\cdot \bv\, ,
\end{equation} 
whose contribution to the collision term is 
\begin{equation}
\frac{1}{2}C^{(2)}_{\Delta v (II)}(\bp)= -\frac{3n_e \sigma_T v}{2p} \int dp' p' \delta(p-p') \int_{-1}^1 
\frac{d\mu'}{2} (f^{(1)}(\bpp)-f^{(1)}(\bp)) (\mu+\mu')\int_0^{2\pi} \frac{d \phi'}{2\pi} 
\cos \theta(1-\cos\theta)\, .
\end{equation}
This integration proceeds through the same steps as for $C^{(2)}_{\Delta v (I)}(\bp)$. In particular
by noting that $\cos \theta(1-\cos \theta)=-1/3+P_1(\cos \theta)-2P_3(\cos\theta)/3$, 
Eqs.~(\ref{DL2}) and~(\ref{intphi}) allows to compute    
\begin{equation}
\int \frac{d \phi'}{2 \pi} \cos \theta (1-\cos \theta)=
-\frac{1}{3} +P_1(\mu) P_1(\mu')-\frac{2}{3} P_2(\mu) P_2(\mu')\, , 
\end{equation} 
and using the decomposition~(\ref{dec1}) we arrive at
\begin{equation}
\label{interm2}
\frac{1}{2}C^{(2)}_{\Delta v(II)}(\bp)=- n_e \sigma_T 
\left\{ {\bf v} \cdot {\bf n}\, f^{(1)}_2(p) (1-P_2({\bf {\hat v}}
\cdot {\bf n}))+ v \left[\frac{1}{5} \,
P_2({\bf {\hat v}}\cdot {\bf n})\,  \left( 
 3 f^{(1)}_1(p)-3 f^{(1)}_3(p) \right)
\right]
\right\}\, .
\end{equation}
Thus summing Eq.~(\ref{interm}) and~(\ref{interm2}) we obtain
\begin{eqnarray}
\frac{1}{2}C^{(2)}_{\Delta v}(\bp)&=&
n_e \sigma_T \Bigg\{ 
{\bf v} \cdot {\bf n} \left[ 
f^{(1)}(\bp) -f^{(1)}_0(p) - p\frac{\partial f^{(1)}_0(p)}{\partial p} 
-f^{(1)}_2(p) 
+ \frac{1}{2} P_2({\bf \hat{v}} \cdot {\bf n}) \left( f^{(1)}_2(p)-
p\frac{\partial f^{(1)}_2(p)}{\partial p}
\right) \right] \nonumber \\
&+& v \left[ 2f^{(1)}_1(p)+ p\frac{\partial f^{(1)}_1(p)}{\partial p} + \frac{1}{5} 
P_2({\bf \hat{v}}\cdot {\bf n}) \left( -f^{(1)}_1(p)+ p\frac{\partial f^{(1)}_1(p)}{\partial p}+
6 f^{(1)}_3(p)+\frac{3}{2} p\frac{\partial f^{(1)}_3(p)}{\partial p}
\right)
\right]
\Bigg\} \, .
\end{eqnarray}
As far as the remaining terms, these have already been computed in Ref.~\cite{DJ} (see also 
Ref.~\cite{HSS})  and here we just report them
\\
\\
d) \quad $c^{(2)}_{v v}(\bp,\bpp)$
\\
\\
The term proportional to the velocity squared yield a contribution to the collision term
\begin{equation}
\frac{1}{2}C^{(2)}_{v v}(\bp)= n_e \sigma_T \left\{ ({\bf v} \cdot {\bf n})^2 \left[ 
p \frac{\partial f^{(0)}}{\partial p}+\frac{11}{20} p^2 
\frac{\partial^2 f^{(0)}}{\partial p^2}
\right]+v^2 \left[ 
p \frac{\partial f^{(0)}}{\partial p}+\frac{3}{20} p^2 
\frac{\partial^2 f^{(0)}}{\partial p^2}
\right]
\right\}\, .
\end{equation}
\\
\\
e) \quad $c^{(2)}_K(\bp,\bp')$
\\
\\
The terms responsible for the spectral distortions give
\begin{equation}
\frac{1}{2}C_K^{(2)}(\bp)=\frac{1}{m_e^2}\frac{\partial}{\partial p}\left\{ 
p^4\left[
T_e \frac{\partial f^{(0)}}{\partial p}+f^{(0)}(1+f^{(0)})
\right]
\right\}\, .
\end{equation}
Finally we write also that part of the collision term coming from Eq.~(\ref{deltac1})
\begin{equation}
\delta^{(1)}_e\, c^{(1)}(\bp,\bpp) \rightarrow \delta^{(1)}_e C^{(1)}(\bp) =
 n_e \sigma_T\, \delta^{(1)}_e \left[ f^{(1)}_0(p)+\frac{1}{2}f^{(1)}_2 P_2({\bf \hat v} \cdot 
{\bf n})-f^{(1)}-p\frac{\partial f^{(0)}}{\partial p} {\bf v} \cdot {\bf n} 
\right]\, . 
\end{equation}

\subsection{Final expression for the collision term}
\noindent 
Summing all the terms we find the final expression for the collision term~(\ref{Integralcolli}) up 
to second order
\begin{eqnarray}
C(\bp)=C^{(1)}(\bp)+\frac{1}{2} C^{(2)}(\bp)
\end{eqnarray}
with 
\begin{eqnarray}
\label{C1p}
C^{(1)}(\bp)= n_e \sigma_T \left[ f^{(1)}_0(p)+\frac{1}{2}f^{(1)}_2 P_2({\bf \hat v} \cdot 
{\bf n})-f^{(1)}-p\frac{\partial f^{(0)}}{\partial p} {\bf v} \cdot {\bf n} \right]
\end{eqnarray}
and 
\begin{eqnarray}
\label{C2p}
\frac{1}{2}C^{(2)}(\bp)&=&n_e \sigma_T \Bigg\{ 
\frac{1}{2}f^{(2)}_{0 0}(p)-\frac{1}{4} \sum_{m=-2}^{2} 
\frac{\sqrt{4 \pi}}{5^{3/2}}\, f^{(2)}_{2m}(p) \, Y_{2m}({\bf n})-\frac{1}{2}
f^{(2)}(\bp)
+\delta^{(1)}_e \left[ f^{(1)}_0(p)+\frac{1}{2}f^{(1)}_2 P_2({\bf \hat v} \cdot 
{\bf n})-f^{(1)}-p\frac{\partial f^{(0)}}{\partial p} {\bf v} \cdot {\bf n} 
\right] \nonumber \\
&-& \frac{1}{2}p\frac{\partial f^{(0)}}{\partial p} {\bf v}^{(2)} 
\cdot {\bf n}+
{\bf v} \cdot {\bf n} \left[ 
f^{(1)}(\bp) -f^{(1)}_0(p) - p\frac{\partial f^{(1)}_0(p)}{\partial p} 
-f^{(1)}_2(p) 
+ \frac{1}{2} P_2({\bf \hat{v}} \cdot {\bf n}) \left( f^{(1)}_2(p)-
p\frac{\partial f^{(1)}_2(p)}{\partial p}
\right) \right] \nonumber \\
&+& v \left[ 2f^{(1)}_1(p)+ p\frac{\partial f^{(1)}_1(p)}{\partial p} + \frac{1}{5} 
P_2({\bf \hat{v}}\cdot {\bf n}) \left( - f^{(1)}_1(p)+ p\frac{\partial f^{(1)}_1(p)}{\partial p}+
6 f^{(1)}_3(p)+\frac{3}{2} p\frac{\partial f^{(1)}_3(p)}{\partial p}
\right)
\right] \nonumber \\
&+& ({\bf v} \cdot {\bf n})^2 \left[ 
p \frac{\partial f^{(0)}}{\partial p}+\frac{11}{20} p^2 
\frac{\partial^2 f^{(0)}}{\partial p^2}
\right]+v^2 \left[ 
p \frac{\partial f^{(0)}}{\partial p}+\frac{3}{20} p^2 
\frac{\partial^2 f^{(0)}}{\partial p^2}
\right]+\frac{1}{m_e^2}\frac{\partial}{\partial p}\left[
p^4\left(
T_e \frac{\partial f^{(0)}}{\partial p}+f^{(0)}(1+f^{(0)})
\right)
\right]
\Bigg\}\, . \nonumber \\
\end{eqnarray}
Notice that there is an internal hierarchy, with terms which do not depend on the baryon velocity 
${\bf v}$, terms proportional to ${\bf v} \cdot {\bf n}$ and then to $({\bf v} \cdot {\bf n})^2$, 
$v$ and $v^2$ (apart from the Kompaneets terms). In particular notice the term  proportional to 
$\delta^{(1)}_e {\bf v} \cdot {\bf n}$ is the one corresponding to the Vishniac effect. 
We point out that we have kept all the 
terms up to second-order in the collision term. In Refs.~\cite{DJ,HSS} many terms coming from $c^{(2)}_{\Delta v}$ 
have been dropped mainly because these terms are proportional to the photon distribution function $f^{(1)}$ which on 
very small scales (those of interest for reionization) is suppressed by the diffusion damping. Here we want 
to be completely general and we have to keep them.   
\\
\\
\section{The Brightness equation}
\noindent 
\subsection{First-order}
\noindent 
The Boltzmann equation for photons is obtained by combining Eq.~(\ref{LHSBoltz}) with Eqs.~(\ref{C1p})-(\ref{C2p}). 
At first-order the left-hand side reads
\begin{eqnarray}
\label{Bdf1}
\frac{df}{d\eta}&=&\frac{d f^{(1)}}{d \eta}-p\frac{\partial f^{(0)}}{\partial p} \frac{\partial \Phi^{(1)}}{\partial x^i} 
\frac{dx^i}{d\eta}+p\frac{\partial f^{(0)}}{\partial p} \frac{\partial \Psi^{(1)}}{ \partial \eta}\, .
\end{eqnarray}
At first-order it is useful to characterize the perturbations to the Bose-Einstein distribution function~(\ref{BEd}) 
in terms of a perturbation to the temperature as 
\begin{equation}
\label{ft}
f(x^i,p,n^i,\eta)=2\, \left[ \exp\left\{ \frac{p}{T(\eta)(1+ \Theta^{(1)})}\right\}-1 \right]^{-1}\, .
\end{equation}  
Thus it turns out that 
\begin{equation}
\label{thetaf1}
f^{(1)}=-p\frac{\partial f^{(0)}}{\partial p} \Theta^{(1)}\, ,
\end{equation} 
where we have used that $\partial f/\partial \Theta|_{\Theta=0}=-p \partial f^{(0)}/\partial p$. In terms of this variable $\Theta^{(1)}$   
the linear collision term~(\ref{C1p}) will now become proportional to $- p \partial f^{(0)}/\partial p$ which contains the only explicit 
dependence on $p$, and the same happens for the left-hand side, Eq.~(\ref{Bdf1}). This is telling us that at first order $\Theta^{(1)}$ 
does not depend on $p$ but only on ${x^i, n^i, \eta}$,  
$\Theta^{(1)}=\Theta^{(1)}({x^i, n^i, \tau})$. This is well known and the physical reason is that at linear order there is no energy 
transfer in Compton collisions between photons and electrons. Therefore the Boltzmann equation for $\Theta^{(1)}$ reads
\begin{equation}
\label{BoltzTheta}
\frac{\partial \Theta^{(1)}}{\partial \eta}+n^i \frac{\partial \Theta^{(1)}}{\partial x^i}
+\frac{\partial \Phi^{(1)}}{\partial x^i} n^i -\frac{\partial \Psi^{(1)}}{ \partial \eta}=
n_e\sigma_T a \left[ \Theta^{(1)}_0+\frac{1}{2} \Theta^{(1)}_2 P_2({\bf {\hat v}} \cdot {\bf n})-\Theta^{(1)}+{\bf v} \cdot {\bf n} 
\right]\, ,
\end{equation} 
where we used that $f^{(1)}_{\ell}=- p \partial f^{(0)}/\partial p \Theta^{(1)}_\ell$ according the decomposition of 
Eq.~(\ref{dec1}), and we have 
taken the zero-order expressions for $dx^i/d\eta$, dropping the contribution from $dn^i/d\eta$ in Eq.~(\ref{Df}) since it is already 
first-order. 

Notice that since $\Theta^{(1)}$ is independent on $p$ it is equivalent to consider the quantity
\begin{eqnarray}
\label{Delta1}
\Delta^{(1)}(x^i,n^i,\tau)&=&
\frac{\int dp p^3 f^{(1)}}{\int dp p^3 f^{(0)}}\, ,\\
\Delta^{(1)}&=&4\Theta^{(1)}\, .
\end{eqnarray}
The physical meaning of $\Delta^{(1)}$ is that of a fractional energy perturbation (in a given direction). From Eq.~(\ref{LHSBoltz}) another 
way to write an equation for $\Delta^{(1)}$ -- the so called brightness equation -- is     
\begin{equation}
\label{B1}
\frac{d}{d\eta} \left[ \Delta^{(1)}+4 \Phi^{(1)} \right]-4 \frac{\partial}{\partial \eta}\left( \Phi^{(1)}+\Psi^{(1)} \right)=
n_e \sigma_T a \left[ \Delta^{(1)}_0+\frac{1}{2} \Delta^{(1)}_2 P_2({\bf {\hat v}} \cdot {\bf n})-\Delta^{(1)}+4 {\bf v} \cdot {\bf n}
\right]\, .
\end{equation}
\subsection{Second-order}
\noindent 
The previous results show that at linear order the photon distribution function has a Planck spectrum with the temperature that at any point 
depends on the photon direction. At second-order one could characterize the perturbed 
photon distribution function in a similar way as in Eq.~(\ref{ft})  
\begin{equation}
f(x^i,p,n^i,\eta)=2\, \left[ \exp\left\{ \frac{p}{T(\eta)\, e^{\Theta}}-1 \right\}\right]^{-1}\, ,
\end{equation}
where by expanding $\Theta=\Theta^{(1)}+\Theta^{(2)}/2+...$ as usual one recovers the first-order expression. 
For example, in terms of $\Theta$, the perturbation of $f^{(1)}$ is given by Eq.~(\ref{thetaf1}), while at second-order  
\begin{eqnarray}
\frac{f^{(2)}}{2}
=-\frac{p}{2} \frac{\partial f^{(0)}}{\partial p} \Theta^{(2)}+\frac{1}{2} \left(p^2 \frac{\partial^2f^{(0)}}{\partial p^2}
+ p \frac{\partial f^{(0)}}{\partial p}  \right) \left( \Theta^{(1)} \right)^2\, .
\end{eqnarray}
However, as discussed in details in Refs.~\cite{HSS,DJ}, now the second-order perturbation $\Theta^{(2)}$ will not be momentum independent 
because the collision term in the equation for $\Theta^{(2)}$ does depend explicitly on $p$ (defining the combination  
$- (p \partial f^{(0)}/\partial p)^{-1} f^{(2)}$ does not lead to a second-order momentum independent equation as above). Such dependence 
is evident, for example, in the terms of $C^{(2)}({\bf p})$, Eq.~(\ref{C2p}), proportional to $v$ or $v^2$, and in the Kompaneets terms. 
The physical reason is that at the non-linear level photons and electrons do exchange energy during Compton collisions. As a consequence 
spectral 
distorsions are generated. For example, in the isotropic limit,
 only the Kompaneets terms survive giving rise to the Sunyaev-Zeldovich distorsions. As discussed in Ref.~\cite{HSS}, the Sunyaev-Zeldovich 
distorsions can also be obtained with the correct coefficients by replacing the average over the direction electron 
$\langle v^2 \rangle$ with the mean squared thermal velocity $\langle v_{th}^2 \rangle=3T_e/m_e$ in Eq.~(\ref{C2p}). 
This is due simply to the fact that 
the distinction between thermal and bulk velocity of the electrons is just for convenience. This fact also 
shows that spectral distorsions due 
to the bulk flow (kinetic Sunyaev-Zeldovich) has the same form as the thermal effect. Thus 
spectral distorsions can be in general described by a global Compton $y$-parameter 
(see Ref.~\cite{HSS} for a full discussion of spectral distorsions). However in the following we will not be 
interested in the frequency dependence but only in the anisotropies of the radiation distribution. Therefore we can integrate over the 
momentum 
$p$ and define ~\cite{HSS,DJ} 
\begin{equation}
\label{Delta2}
\Delta^{(2)}(x^i,n^i,\tau)=\frac{\int dp p^3 f^{(2)}}{\int dp p^3 f^{(0)}}\, ,
\end{equation}
as in Eq.~(\ref{Delta1}).

Integration over $p$ of Eqs.~(\ref{LHSBoltz})-(\ref{C2p}) is straightforward using the following relations
\begin{eqnarray}
\label{rules}
\int dp p^3 p \frac{\partial f^{(0)}}{\partial p}&=&-4\, N\, , \nonumber\\
\int dp p^3 p^2 \frac{\partial^2 f^{(0)}}{\partial p^2}&=& 20\, N\, , \nonumber\\
\int dp p^3 f^{(1)} &=&N \Delta^{(1)} \, ,\nonumber \\
\int dp p^3 p \frac{\partial f^{(1)}}{\partial p}&=&-4\, N \Delta^{(1)}\, ,
\nonumber\\
N&=&\int dp p^3 f^{(0)}\, .
\end{eqnarray}
where $N$ is the normalization factor in Eq.~(\ref{Delta2}) 
(it is just proportional the background energy density of 
photons ${\bar \rho}_\gamma$). At first-order one recovers Eq.~(\ref{B1}). At second-order we find
\begin{eqnarray}
\label{B2}
& & \frac{1}{2} \frac{d}{d\eta} \left[ \Delta^{(2)}+4\Phi^{(2)} \right] + \frac{d}{d\eta} \left[ \Delta^{(1)} +4 \Phi^{(1)} \right] 
-4 \Delta^{(1)}\left( \Psi^{(1)'}-\Phi^{(1)}_{,i}n^i \right) -2 \frac{\partial}{\partial \eta}\left( \Psi^{(2)}+\Phi^{(2)} \right) 
+4 \frac{\partial \omega_i}{\partial \eta} n^i + 2 \frac{\partial \chi_{ij}}{\partial \eta} n^i n^j \nonumber \\
& &= - \frac{\tau'}{2} \Bigg[ \Delta^{(2)}_{00} -\Delta^{(2)} 
- \frac{1}{2} \sum_{m=-2}^{2} \frac{\sqrt{4 \pi}}{5^{3/2}}\, \Delta^{(2)}_{2m} \, Y_{2m}({\bf n}) +
2 \delta^{(1)}_e \left( \Delta^{(1)}_0+\frac{1}{2} \Delta^{(1)}_2 P_2({\bf {\hat v}} \cdot {\bf n})-\Delta^{(1)}+4 {\bf v} \cdot {\bf n}
\right)+4{\bf v}^{(2)} \cdot {\bf n} \nonumber \\
& & + 2 ({\bf v} \cdot {\bf n}) \left[ \Delta^{(1)}+3\Delta^{(1)}_0-\Delta^{(1)}_2 \left(1-\frac{5}{2} P_2({\bf {\hat v}} 
\cdot {\bf n})
\right)\right]-v\Delta^{(1)}_1 \left(4+2 P_2({\bf {\hat v}} \cdot {\bf n}) \right) 
+14 ({\bf v} \cdot {\bf n})^2-2 v^2  \Bigg]\, ,
\end{eqnarray}
where we have expanded the angular dependence of $\Delta$ as in Eq.~(\ref{fangdeco})
\begin{equation}
\label{Dlm2}
\Delta^{(i)}({\bf x}, {\bf n})=\sum_{\ell} \sum_{m=-\ell}^{\ell} \Delta^{(i)}_{\ell m}({\bf x})  
(-i)^{\ell}  \sqrt{\frac{4 \pi}{2\ell+1}}Y_{\ell m}({\bf n})\, ,
\end{equation} 
with  
\begin{equation}
\label{angular1}
\Delta^{(i)}_{\ell m}=(-i)^{- \ell}\sqrt{\frac{2\ell+1}{4\pi}} \int d\Omega  \Delta^{(i)} 
Y^{*}_{\ell m}({\bf n}) \, ,
\end{equation}
where we recall that the superscript stands by the order of the perturbation. At first order one can drop the dependence on $m$ 
setting $m=0$ so that  $\Delta^{(1)}_{\ell m}=(-i)^{-\ell} (2\ell +1)  \delta_{m0} \, \Delta^{(1)}_{\ell}$. In Eq.~(\ref{B2}) we have 
introduced the differential optical depth 
\begin{equation}
\label{deftau}
\tau'=-{\bar n}_e \sigma_T a \, .
\end{equation}
It is understood that
 on the left-hand side of Eq.~(\ref{B2}) one has to 
pick up for the total time derivatives only those terms which contribute to 
second-order. Thus we have to take 
\begin{eqnarray}
\frac{1}{2} \frac{d}{d\eta} \left[ \Delta^{(2)}+4\Phi^{(2)} \right] + \frac{d}{d\eta} \left[ \Delta^{(1)} +4 \Phi^{(1)} \right]\Big|^{(2)}
&=& \frac{1}{2}\left( \frac{\partial}{\partial \eta}+n^i \frac{\partial}{\partial x^i}\right) \left( \Delta^{(2)}+4\Phi^{(2)} \right)
+n^i(\Phi^{(1)}+\Psi^{(1)})\partial_i(\Delta^{(1)}+4\Phi^{(1)}) \nonumber\\ 
&+& \left[(\Phi^{(1)}_{,j}+\Psi^{(1)}_{,j})n^in^j 
-(\Phi^{,i}+\Psi^{,i})\right]
\frac{\partial \Delta^{(1)}}{\partial n^i}\, , 
\end{eqnarray}
where we used Eqs.~(\ref{dxi}) and ~(\ref{dni}). Notice that we can write $\partial \Delta^{(1)}/\partial n^i=
(\partial \Delta^{(1)}/\partial x^i) (\partial x^i /\partial n^i)=(\partial \Delta^{(1)}/\partial x^i) (\eta-\eta_i)$, from the 
integration in time of Eq.~(\ref{dxi}) at zero-order when $n^i$ is constant in time. 

\subsection{Hierarchy equations for multipole moments}
\noindent 
Let us now move to Fourier space. In the following for a given ${\bf k}$ mode we will choose the coordinate system such that 
${\bf e}_3={\bf {\hat k}}$ and the polar angle 
of the photon momentum is $\vartheta$, with $\mu=\cos \vartheta = {\bf \hat{k}} \cdot {\bf n}$. Then Eq.~(\ref{B2}) can be written as 
\begin{equation}
\label{BF}
\Delta^{(2) \prime}+ik\mu \Delta^{(2)}-\tau' \Delta^{(2)}= S({\bf k},{\bf n},\eta)\, ,
\end{equation}
where $S({\bf k},{\bf n},\eta)$ can be easily read off Eq.~(\ref{B2}). 
We now expand the temperature anisotropy in the multipole moments $\Delta^{(2)}_{\ell m}$ in order to obtain a system of coupled 
differential equations. By applying the angular integral of Eq.~(\ref{angular1}) to Eq.~(\ref{BF}) we find
\begin{equation}
\label{H}
\Delta^{(2)\prime }_{\ell m}({\bf k},\eta)=k \left[\frac{{\kappa}_{\ell m}}{2\ell-1}\Delta^{(2)}_{\ell-1,m}-
\frac{{\kappa}_{\ell+1, m}}{2\ell+3}\Delta^{(2)}_{\ell+1,m} \right] + \tau' \Delta^{(2)}_{\ell m}+ S_{\ell m}\, ,
\end{equation}  
where the expansion coefficients of the source term are given by
\begin{eqnarray}
\label{Slm}
S_{\ell m}&=& \left( 4 \Psi^{(2)\prime }-\tau' \Delta^{(2)}_{00} \right) \delta_{\ell0} \delta_{m0} +4 k \Phi^{(2)} 
\delta_{\ell 1} \delta_{m 0}-4\, \omega_{\pm 1} \delta_{\ell 1} -8 \tau' v^{(2)}_m \delta_{\ell 1}-\frac{\tau'}{10} 
\Delta^{(2)}_{\ell m} \,\delta_{\ell 2} -2\chi_{\pm 2} \,\delta_{\ell 2} \nonumber \\
&-&2 \tau' \int \frac{d^3k_1}{(2\pi)^3} \left[ v^{(1)}_0({\bf k}_1) v^{(1)}_0({\bf k}_2) {\bf {\hat k}}_2 \cdot  
{\bf {\hat k}}_1  + \delta^{(1)}_e({\bf k}_1) \Delta^{(1)}_0({\bf k}_2) -i\frac{2}{3} 
v({\bf k}_1) \Delta^{(1)}_{10}({\bf k}_2) \right] \delta_{\ell 0} \delta_{m 0}\nonumber \\
&+& k\, \int \frac{d^3k_1}{(2\pi)^3} \left[ \Phi^{(1)}({\bf k}_1) \Phi^{(1)}({\bf k}_2) 
\right] \delta_{\ell 1} \delta_{m 0} -2 \left[ (\Psi^{(1)} \nabla \Phi^{(1)})_m +8\tau' (\delta^{(1)}_e {\bf v})_m + +6 \tau' 
(\Delta^{(1)}_0 {\bf v})_m-2 \tau' (\Delta^{(1)}_2 {\bf v})_m  \right]\, \delta_{\ell 1}\nonumber \\
&+& \tau'\int \frac{d^3k_1}{(2\pi)^3} \left[ \delta^{(1)}_e({\bf k}_1) \Delta^{(1)}_2({\bf k}_2)
-i\frac{2}{3}v({\bf k}_1) \Delta^{(1)}_{10}({\bf k}_2) \right]\, \delta_{\ell 2} \, \delta_{m 0}
+\int \frac{d^3k_1}{(2\pi)^3} \left[ 8 \Psi^{(1)}({\bf k}_1)+2 \tau' \delta^{(1)}_e({\bf k}_1) \right. \nonumber \\ 
&-& \left. (\eta-\eta_i) (\Psi^{(1)}+\Phi^{(1)})({\bf k}_1) \, {\bf 
k}_1\cdot{\bf k}_2  \right] \Delta^{(1)}_{\ell0}({\bf k}_2)
\delta_{m0}\nonumber \\
&-&i (-i)^{-\ell}(-1)^{-m} (2\ell +1) \sum_{\ell''} \sum_{m'=-1}^{1}(2\ell''+1) \left[ 8 \Delta^{(1)}_{\ell''} \nabla \Phi^{(1)}
-2(\Phi^{(1)}+\Psi^{(1)}) \nabla \Delta^{(1)}_{\ell''}\right]_{m'}
\left(\begin{array}{ccc}\ell''&1&\ell\\0&0&0\end{array}\right)
\left(\begin{array}{ccc}\ell''&1&\ell\\0&m'&-m\end{array}\right) \nonumber \\
&+&\tau' i (-i)^{-\ell}(-1)^{-m} (2\ell +1) \sum_{\ell''} \sum_{m'=-1}^{1}(2\ell''+1) \left[ 
2 \Delta^{(1)}_{\ell''} {\bf v} +5   \delta_{\ell'' 2} \, \Delta^{(1)}_2 {\bf v} \right]_{m'}
\left(\begin{array}{ccc}\ell''&1&\ell\\0&0&0\end{array}\right)
\left(\begin{array}{ccc}\ell''&1&\ell\\0&m'&-m\end{array}\right) \nonumber \\
&+&14\tau'(-i)^{-\ell}(-1)^{-m} \sum_{m',m''=-1}^{1} 
\int \frac{d^3k_1}{(2\pi)^3} \left[ v^{(1)}_0({\bf k}_1) \frac{v^{(1)}_0({\bf k}_2)}{k_2} \frac{4\pi}{3} 
Y^*_{1m'}({\bf {\hat k}}_1) \left( k Y^*_{1m''}({\bf {\hat k}})- k_1 Y^*_{1m''}({\bf {\hat k}_1})\right)
\right]  \nonumber \\ 
&\times&\left(\begin{array}{ccc}1&1&\ell\\0&0&0\end{array}\right)
\left(\begin{array}{ccc}1&1&\ell\\m'&m''&-m\end{array}\right) \nonumber \\
&+&2 (\eta-\eta_i)(-i)^{-\ell} (-1)^{-m} \sqrt{\frac{2\ell+1}{4\pi}} \sum_L \sum_{m',m''=-1}^{1} 
\int \frac{d^3k_1}{(2\pi)^3} \sqrt{\frac{4\pi}{2L+1}} \left( \frac{4\pi}{3} \right)^2 
\Delta^{(1)}_L({\bf k}_1) (\Phi^{(1)}+\Psi^{(1)})({\bf k}_2) \nonumber \\
&\times& k_1 Y^*_{1m'}({\bf {\hat k}}_1)\left( k Y^*_{1m''}({\bf {\hat k}})- k_1 Y^*_{1m''}({\bf {\hat k}_1})\right)
 \int d \Omega Y_{1m'}({\bf n}) Y_{1m''}({\bf n}) Y_{L0}({\bf n}) Y_{\ell-m}({\bf n})\, ,
\end{eqnarray}
where 
\begin{equation}
{\bf k}_2={\bf k}-{\bf k}_1\, .
\end{equation}
and $k_2=|{\bf k}_2|$. In Eq.~(\ref{Slm}) it is understood that $|m| \leq \ell$.

Let us explain the notations we have adopted in writing Eq.~(\ref{Slm}). The baryon velocity at linear order is irrotational, 
meaning that it is the gradient of a potential, and thus in Fourier space it is parallel to 
${\bf {\hat k}}$, and following the 
conventions of Ref.~\cite{husolo}, we write 
\begin{equation}
\label{vzero}
{\bf v}^{(1)}({\bf k})=-i v^{(1)}_0({\bf k}) {\bf {\hat k}}\, .
\end{equation}
For the second-order velocity perturbation, it will contain a transverse (divergence-free) part whose components are orthogonal to  
${\bf {\hat k}}={\bf e}_3$, and we can write 
\begin{equation}
\label{vdec}
{\bf v}^{(2)}({\bf k})=-i v^{(2)}_0({\bf k}) {\bf e}_3+\sum_{m=\pm1} v^{(2)}_m \, \frac{{\bf e}_2\mp i {\bf e}_1}{\sqrt 2}\, ,
\end{equation}
where ${\bf e}_i$ form an orhtonormal basis with  ${\bf {\hat k}}$. The second-order
perturbation $\omega_i$ is decomposed in a similar way, and $\omega_{\pm 1}$ are the corresponding components 
(in this case in the Poisson gauge there is no scalar component). Similarly for the tensor perturbation $\chi_{ij}$ 
we have indicated its amplitudes as $\chi_{\pm 2}$ in the decomposition~\cite{huwhite}
\begin{equation}
\chi_{ij}=\sum_{m=\pm 2} - \sqrt{\frac{3}{8}} \, \chi_m ({\bf e}_1\pm i {\bf e}_2)_i ({\bf e}_1\pm i {\bf e}_2)_j\, .  
\end{equation}
We have taken into account that in the gravitational part of the Boltzmann equation and in the collsion term there are some terms, like 
$\delta^{(1)}_e {\bf v}$, which still can be decomposed in the scalar and transverse parts in Fourier space as in Eq.~(\ref{vdec}). 
For a generic quantity  $f({\bf x}) {\bf v}$ we have indicated the corresponding scalar and vortical components with $(f {\bf v})_m$ and 
their explicit expression is easily found by projecting the Fourier modes of $f({\bf x}) {\bf v}$ along the ${\bf {\hat k}}={\bf e}_3$ and 
$({\bf e}_2\mp i {\bf e}_1)$ directions 
\begin{equation}
(f {\bf v})_m({\bf k})=\int \frac{d^3k_1}{(2\pi)^3} v^{(1)}_0({\bf k}_1) f({\bf k}_2) Y^*_{1m}({\bf \hat{k}}_1) \sqrt{\frac{4\pi}{3}}\, .
\end{equation} 
Similarly for a term like $f({\bf x}) \nabla g({\bf x})$ we used the notation
\begin{equation}
(f\nabla g)_m({\bf k})=- \int \frac{d^3k_1}{(2\pi)^3} k_1 g({\bf k}_1) f({\bf k}_2) Y^*_{1m}({\bf \hat{k}}_1) 
\sqrt{\frac{4\pi}{3}}\, .
\end{equation}   
Finally, the first term on the right-hand side of Eq.~(\ref{H}) has been obtained by using the relation 
\begin{equation}
\label{gradient}
i {\bf k} \cdot {\bf n}\, \Delta^{(2)}({\bf k})= \sum_{\ell m} \Delta^{(2)}_{\ell m}({\bf k}) \frac{k}{2\ell +1} \left[ 
{\kappa_{\ell m} {\tilde G}_{\ell-1,m}-\kappa_{\ell+1,m}{\tilde G}_{\ell+1,m}} \right] =  
k \sum_{\ell m} \left[\frac{{\kappa}_{\ell m}}{2\ell-1}\Delta^{(2)}_{\ell-1,m}-
\frac{{\kappa}_{\ell m}}{2\ell+3}\Delta^{(2)}_{\ell+1,m} \right] {\tilde G}_{\ell m}\, , 
\end{equation} 
where $ {\tilde G}_{\ell m}= (-i)^{\ell} \sqrt{4\pi/(2\ell+1)}Y_{\ell m}({\bf n})$ is the 
angular mode for the decomposition~(\ref{Dlm2}) and 
\begin{equation}
{\kappa}_{\ell m}=\sqrt{l^2-m^2}\, .
\end{equation} 
This relation has been discussed in Refs.~\cite{Complete,huwhite} and corresponds to the term $n^i \partial \Delta^{(2)}/\partial x^i$ 
in Eq.~(\ref{B2}).

As expected at second-order we recover some intrinsic effects which are characteristic of the linear regime. 
In Eq.~(\ref{H}) the relation~(\ref{gradient})  represents the free streeming effect: 
when the radiation free streems the  inhomogeneities of the photon    
distribution are seen by the obsever as angular anisotropies. At first order it is responsible for the hierarchy of Boltzmann 
equations coupling the different $\ell$ modes, and it represents a projection effect of fluctuations on a scale $k$ onto the angular scale 
$\ell \sim k\eta$. The term $\tau' \Delta^{(2)}_{\ell m}$ causes an exponential suppression of anisotropies in the absence of the source 
term $S_{\ell m}$. The first line of the source term~(\ref{Slm}) just reproduces the expression of the first order case.   
Of course the dynamics of the second-order metric and baryon-velocity perturbations which appear will be different and governed by the 
second-order Einstein equations and continuity equations. The remaining terms in the source are second-order effects generated as 
non-linear combinations of the primordial (first-order) perturbations. We have ordered them according to the increasing number of 
$\ell$ modes they contribute to. Notice in particular that they involve the first-order anisotropies $\Delta^{(1)}_\ell$  and as a 
consequence such terms contribute to generate the hierarchy of equations (apart from the free streeming effect). The source term 
contains additional scattering processes and gravitational effects. On large scales (above the horizon at recombination) 
we can say that the main effects are due to gravity, and they include the Sachs-Wolfe and the (late and early) Sachs-Wolfe effect due 
to the redshift photons suffer when travelling through the second-order gravitational potentials. These, toghether with the contribution 
due to the second-order tensor modes, have been already studied in details in Ref.~\cite{fulllarge}. 
Another important gravitational effect is 
that of lensing of photons as they travel from the last scattering surface to us. A contribution of this type is given by the last term
of Eq.~(\ref{Slm}). On small scales scattering effects around the epoch of recombination are important and we will study them in 
details in a companion paper~\cite{toappear}.

\subsection{Integral solution to the second-order Boltzmann equation}
\noindent 
As in linear theory, one can derive an integral solution to the Boltzmann equation~(\ref{B2}) in terms of the source term $S$.  
Following the standard procedure (see {\it e.g.} Ref.~\cite{Dodelsonbook,SeZ}) for linear perturbations, 
we write the left-hand side as $\Delta^{(2) \prime}+ik\mu \Delta^{(2)}-\tau' \Delta^{(2)}=
e^{-ik\mu+\tau} d[\Delta^{(2)} e^{ik \mu \eta-\tau}]/d\eta$ in order to derive the integral solution  
\begin{eqnarray}
\label{IS}
\Delta^{(2)}({\bf k},{\bf n},\eta_0)=\int_0^{\eta_0} d\eta S({\bf k},{\bf n},\eta) e^{ik\mu(\eta-\eta_0)-\tau}\, ,
\end{eqnarray}
where $\eta_0$ stands by the present time. The expression of the photon moments $\Delta^{(2)}_{\ell m}$ can be 
obtained as usual from Eq.~(\ref{angular1}). In the previous section we have already found the coefficients for the 
decomposition of source term $S$  
\begin{eqnarray}
\label{decS}
S({\bf k},{\bf n}, \eta)= \sum_{\ell} \sum_{m=-\ell}^{\ell} S_{\ell m}({\bf k},\eta) 
(-i)^{\ell}\ \sqrt{\frac{4\pi}{2\ell+1}} Y_{\ell m}({\bf n}) \, .
\end{eqnarray}
In Eq.~(\ref{IS}) there is an additional angular dependence in the exponential. It is easy to 
take it into account by recalling that 
\begin{equation}
\label{eikx}
e^{i{\bf k} \cdot {\bf x}}=
\sum_\ell (i)^\ell (2\ell+1) j_\ell(kx) P_\ell( {\bf {\hat k}} \cdot {\bf {\hat x}}) \, .
\end{equation}
Thus the angular integral~(\ref{angular1}) is computed by using the decomposition of the source term~(\ref{decS}) 
and Eq.~(\ref{eikx}) 
\begin{eqnarray}
\label{IS1}
\Delta^{(2)}_{\ell m}( {\bf k},\eta_0)&=&(-1)^{-m}(-i)^{-\ell} (2\ell+1) \int_0^{\eta_0} d\eta \, e^{-\tau(\eta)}  
\sum_{\ell_2}\sum_{m_2=-\ell_2}^{\ell_2} (-i)^{\ell_2} S_{\ell_2 m_2}  \sum_{\ell_1} i^{\ell_1} 
(2\ell_1+1) j_{\ell_1}(k(\eta-\eta_0)) (2\ell_1+1)  
\nonumber \\  
&\times& \left(\begin{array}{ccc}\ell_1&\ell_2&\ell\\0&0&0\end{array}\right)
\left(\begin{array}{ccc}\ell_1&\ell_2&\ell\\0&m_2&- m\end{array}\right) \, ,
\end{eqnarray}
where the Wigner $3-j$ symbols appear because of the Gaunt integrals
\begin{eqnarray}
  \nonumber
  {\mathcal G}_{l_1l_2l_3}^{m_1m_2m_3}
  &\equiv&
  \int d^2\hat{\mathbf n}
  Y_{l_1m_1}(\hat{\mathbf n})
  Y_{l_2m_2}(\hat{\mathbf n})
  Y_{l_3m_3}(\hat{\mathbf n})\\
 \nonumber
  &=&\sqrt{
   \frac{\left(2l_1+1\right)\left(2l_2+1\right)\left(2l_3+1\right)}
        {4\pi}
        }
  \left(
  \begin{array}{ccc}
  \ell_1 & \ell_2 & \ell_3 \\ 0 & 0 & 0 
  \end{array}
  \right)
  \left(
  \begin{array}{ccc}
  \ell_1 & \ell_2 & \ell_3 \\ m_1 & m_2 & m_3 
  \end{array}
  \right),
\end{eqnarray}
Since the second of the Wigner 3-$j$ symbols in Eq.~(\ref{IS1}) 
is nonzero only if $m=m_2$, our solution can be rewritten to 
recover the corresponding expression found for linear anisotropies in Refs.~\cite{huwhite,Complete}
\begin{equation}
\label{comp}
\frac{\Delta^{(2)}_{\ell m}( {\bf k},\eta_0)}{2\ell+1}=
\int_0^{\eta_0} d\eta \, e^{-\tau(\eta)} \sum_{\ell_2}\sum_{m_2=-\ell_2}^{\ell_2} S_{\ell_2 m_2} j_{\ell}^{(l_2,m)}[k(\eta_0-\eta)]\, ,
\end{equation}    
where $j_{\ell}^{(l_2,m)}[k(\eta_0-\eta)]$ are the so called radial functions. Of course the main information at second-order is 
included in the source term containing different effects due to the non-linearity of the perturbations. 
In the total angular momentum method of Refs.~\cite{huwhite,Complete} Eq.~(\ref{comp}) is interpreted just as the intergration over 
the radial coordinate $(\chi=\eta_0-\eta)$ of the projected source term. 
Another important comment is that, as in linear theory, the integral solution~(\ref{IS1}) is in fact just a formal solution, since 
the source term $S$ contains itself the second-order photon moments up to $l=2$ (see Eq.~(\ref{Slm})). 
This means that one has anyway to resort to the 
hierarchy equations for photons, Eq.~(\ref{H}), to solve for these moments. Neverthless, as in linear theory~\cite{SeZ}, 
one expects to need just a few moments beyond $\ell=2$ in the hierarchy equations, and once the moments entering in the 
source function are computed the higher moments are obtained from the integral solution. Thus the integral solution should in fact be 
more advantageous than solving the system of coupled equations~(\ref{H}).

\section{The Boltzmann equation for baryons and CDM}
\noindent 
In this section we will derive the Boltzmann equation for massive particles, 
which is the case of interest for baryons and dark matter. These equations are necessary to find the time evolution of number 
densities and velocities of the baryon fluid which appear in the brightness equation, thus allowing to close the system of 
equations. Let us start from the baryon component. 
Electrons are tightly coupled to protons via Coulomb interactions. This forces the relative energy density contrasts and the 
velocities to a common value, $\delta_e=\delta_p \equiv \delta_b$ and ${\bf v}_e={\bf v}_p \equiv {\bf v}$, so that we can identify  
electrons and protons collectively as ``baryonic'' matter. 

To derive the Boltzmann equation for baryons let us first focus 
on the collisionless equation and compute therefore $dg/d\eta$, where $g$ is the distribution function for a massive 
species with mass $m$. One of the differences with respect to photons is just that baryons are non-relativistic for the epochs of 
interest. Thus the first step is to generalize the formulae in Section II up to Eq.~(\ref{dni}) to the case of a massive particle. 
In this case one enforces the constraint $Q^2=g_{\mu\nu}Q^\mu Q^\nu=-m^2$ and it also useful to use the particle energy
\begin{equation}
E=\sqrt{q^2+m^2}\, ,
\end{equation}     
where $q$ is defined as in Eq.~(\ref{defp}). Moreover in this case it is very convenient to take 
the distribution function as a function of the variables $q^i=qn^i$, the 
position $x^i$ and time $\eta$, without using the explicit splitting into the magnitude of the momentum $q$ (or the energy E) and its direction $n^i$. Thus the total time derivative of the distribution functions reads
\begin{equation}
\label{DG}
\frac{d g}{d \eta}=\frac{\partial g}{\partial \eta}+
\frac{\partial g}{\partial x^i} \frac{d x^i}{d \eta}+
\frac{\partial g}{\partial q^i} \frac{d q^i}{d \eta}\, .
\end{equation}
We will not give the details of the calculation since we just need to replicate the 
same computation we did for the photons. For the four-momentum of the particle notice that $Q^i$ has the same form as Eq.~(\ref{Pi}), 
while for $Q^0$ we find
\begin{equation}
\label{P0m}
Q^0=\frac{e^{-\Phi}}{a}\, E \left(1+\omega_i \frac{q^i}{E}  \right)\, .
\end{equation}  
In the following we give the expressions for $dx^i/d\eta$ and $dq^i/d\eta$.
\\
\\
a) As in Eq.~(\ref{dxi}) $dx^i/d\eta=Q^i/Q^0$ and it turns out to be
\begin{equation}
\frac{d x^i}{d\eta}=\frac{q}{E} n^i e^{\Phi+\Psi}  \left(1-\omega_i n^i \frac{q}{E} \right) 
\left(1-\frac{1}{2} \chi_{km} n^k n^m \right)\, .
\end{equation}
\\
\\
b) For $dq^i/d\eta$ we need the expression of $Q^i$ which is the same as that of  
Eq.~(\ref{Pi}) 
\begin{equation}
\label{Qi}
Q^i= \frac{q^i}{a} e^{\Psi} \left( 1-\frac{1}{2} \chi_{km}n^kn^m \right)\, .
\end{equation}
The spatial component of the geodesic equation up to second-order reads 
\begin{eqnarray}
\label{SPG}
\frac{dQ^i}{d\eta}&=&-2({\cal H}-\Psi')\left( 1-\frac{1}{2}\chi_{km}n^kn^m \right) 
\frac{q}{a}n^i e^{\Psi}+2 \frac{\partial \Psi}{\partial x^k}\frac{q^2}{aE}n^in^ke^{\Phi+2\Psi}
-\frac{\partial \Phi}{\partial x^i}\frac{E}{a}e^{\Phi+2\Psi}-
\frac{\partial \Psi}{\partial x^i}\frac{q^2}{aE}e^{\Phi+2\Psi} 
-(\omega^{i\prime}+{\cal H} \omega^i) \frac{E}{a} \nonumber \\
&-&(\chi^{i\prime}_{~k}+\omega^{i}_{'k}-\omega_k^{,i})\frac{E}{a} 
+\left[{\cal H}\omega^i \delta_{jk}-(\chi^{i}_{~j,k}+\chi^i_{~k,j}+\chi_{jk}^{~~,i})\right] \frac{q^j q^k}{Ea}\, .
\end{eqnarray}
Proceeding as in the massless case we now take the total time derivative of Eq.~(\ref{Qi}) and using Eq.~(\ref{SPG}) we find 
\begin{eqnarray}
\frac{dq^i}{d\eta}&=&-({\cal H}-\Psi')q^i+\Psi_{,k}\frac{q^iq^k}{E} e^{\Phi+\Psi}-\Phi^{,i}E e^{\Phi+\Psi} 
-\Psi_{,i}\frac{q^2}{E} e^{\Phi+\Psi} -E(\omega^{i\prime}+{\cal H} \omega^i)
- (\chi^{i\prime}_{~k}+\omega^{i}_{'k}-\omega_k^{,i}) E\nonumber \\
&+&\left[{\cal H}\omega^i \delta_{jk}-(\chi^{i}_{~j,k}+\chi^i_{~k,j}+\chi_{jk}^{~,i}) \right] \frac{q^j q^k}{E}\, .
\end{eqnarray}
We can now write the total time derivative of the distribution function as
\begin{eqnarray}
\label{Dg}
\frac{d g}{d \eta}&=& 
 \frac{\partial g}{\partial \eta}+\frac{q}{E} n^i e^{\Phi+\Psi}\left(1-\omega_i n^i -\frac{1}{2} 
\chi_{km}n^kn^m \right) \frac{\partial g}{\partial x^i}+
 \left[ -({\cal H}-\Psi')q^i+\Psi_{,k}\frac{q^iq^k}{E} e^{\Phi+\Psi}-\Phi^{,i}E e^{\Phi+\Psi} 
-\Psi_{,i}\frac{q^2}{E} e^{\Phi+\Psi} \right. \nonumber \\
&-& \left. E(\omega^{i\prime}+{\cal H} \omega^i)
- (\chi^{i\prime}_{~k}+\omega^{i}_{'k}-\omega_k^{,i}) E 
+\left({\cal H}\omega^i \delta_{jk}-(\chi^{i}_{~j,k}+\chi^i_{~k,j}+\chi_{jk}^{~,i})\right) \frac{q^j q^k}{E} 
\right]
\frac{\partial g}{\partial q^i}\, .
\end{eqnarray}
This equation is completely general since we have just solved for the kinematics of massive particles. 
As far as the collision terms are concerned, for the system of electrons and protons we consider the Coulomb scattering processes 
between the electrons and protons and the Compton scatterings between photons and electrons 
\begin{eqnarray}
\label{boltzep}
\frac{dg_e}{d\eta}({\bf x},{\bf q},\eta)&=&\langle c_{ep} \rangle_{QQ'q'} +\langle c_{e\gamma} \rangle_{pp'q'} \\
\label{boltzep2}
\frac{dg_p}{d\eta}({\bf x},{\bf Q},\eta)&=&\langle c_{ep} \rangle_{qq'Q'}\, ,
\end{eqnarray} 
where we have adopted the same formalism of Ref.~\cite{Dodelsonbook} with ${\bf p}$ and ${\bf p}'$ the initial and final momenta of the 
photons, ${\bf q}$ and ${\bf q}'$ the corresponding quantities for the electrons and for protons ${\bf Q}$ and ${\bf Q}'$. The 
integral  over different momenta is indicated by   
\begin{equation}
\langle \cdots \rangle_{pp'q'} \equiv \int \frac{d^3p}{(2\pi)^3}\,\int \frac{d^3 p'}{(2\pi)^3}\, 
\int \frac{d^3q'}{(2\pi)^3} \dots \, ,
\end{equation}
and thus one can read 
$c_{e\gamma}$ as the unintegrated part of Eq.~(\ref{collisionterm}), and similarly for $c_{ep}$ (with the appropriate amplitude
$|M|^2$). In Eq.~(\ref{boltzep}) Compton scatterings between protons and photons can be safely neglected because the amplitude of 
this process has a much smaller amplitude than Compton scatterings with electrons being weighted by the inverse squared mass of the 
particles. 

At this point for the photons we considered the perturbations around the zero-order Bose-Einstein distribution function (which are 
the unknown quantities). For the electrons (and protons) we can take the thermal distribution described by Eq.~(\ref{gel}). 
Moreover we will take the moments of Eqs.~(\ref{boltzep})-(\ref{boltzep2}) in order to find the energy-momentum continuity equations.

\subsection{Energy continuity equation} 
\noindent 
We now integrate Eq.~(\ref{Dg}) over $d^3q/(2\pi)^3$. Let us recall that in terms of the distribution function 
the number density $n_e$ and the bulk velocity ${\bf v}$ are given by
\begin{equation}
\label{defne}
n_e=\int \frac{d^3 q}{(2\pi)^3}\, g \, , 
\end{equation} 
and 
\begin{equation}
\label{defvg}
v^i= \frac{1}{n_e} \int \frac{d^3q}{(2\pi)^3}\, g \, \frac{q n^i}{E}\, ,
\end{equation}
where one can set $E\simeq m_e$ since we are considering non-relativistic particles. 
We will also make use of the following relations when integrating over the solid angle $d\Omega$
\begin{equation}
\label{relOmega}
\int d\Omega\, n^i=\int d\Omega\, n^in^jn^k=0\, ,\quad \int \frac{d\Omega}{4\pi}\, n^in^j =  \frac{1}{3} \delta^{ij}\, . 
\end{equation}
Finally notice that $dE/dq=q/E$ and $\partial g/\partial q= (q/E) \partial g/\partial E$.

Thus the first two integrals just brings $n'_e$ and $(n_e v^i)_{,i}$. Notice that all the terms proportional to the second-order 
vector and tensor perturbations of the metric give a vanishing contribution at second-order since in this case we can take the 
zero-order distribution functions which depends only on $\eta$ and $E$, integrate over the direction 
and use the fact that $\delta^{ij} \chi_{ij}=0$. The trick to solve the remaining integrals is an integration  by parts over $q^i$.  
We have an integral like (the one multiplying $( \Psi'-{\cal H})$)
\begin{equation}
\label{r1}
\int \frac{d^3q}{(2 \pi)^3} q^i \frac{\partial g}{\partial q^i} = -3 \int \frac{d^3q}{(2 \pi)^3} g =-3 n_e \, ,
\end{equation}
after an integration by parts over $q^i$. The remaining integrals can be solved still by integrating by parts over $q^i$.  
The integral proportional to $\Phi^{,i}$ in Eq.~(\ref{Dg}) gives  
\begin{equation}
-e^{\Phi+\Psi} \Phi^{,i} \int \frac{d^3q}{(2 \pi)^3} E \frac{\partial g}{\partial q^i}=e^{\Phi+\Psi} \Phi^{,i}v_i\, ,
\end{equation}
where we have used that $dE/dq^i=q^i/E$. For the integral 
\begin{equation}
e^{\Phi+\Psi} \Psi_{,k}\int \frac{d^3q}{(2 \pi)^3} \frac{q^iq^k}{E} \frac{\partial g}{\partial q^i}\, , 
\end{equation}
the integration by parts brings two pieces, one from the derivation of $q^iq^k$ and one from the derivation of the energy $E$
\begin{equation}
\label{exint}
- 4 e^{\Phi+\Psi} \Psi_{,k}\int \frac{d^3q}{(2 \pi)^3} g\frac{q^k}{E} + e^{\Phi+\Psi} \Psi_{,k}    
\int \frac{d^3q}{(2 \pi)^3} g \frac{q^2}{E} \frac{q^k}{E}=-4e^{\Phi+\Psi} \Psi_{,k}v^k+ e^{\Phi+\Psi} \Psi_{,k}    
\int \frac{d^3q}{(2 \pi)^3} g \frac{q^2}{E^2} \frac{q^k}{E} \, .
\end{equation}
The last integral in Eq.~(\ref{exint}) can indeed be neglected. To check this one makes use of the explicit  expression~(\ref{gel}) for the 
distribution function $g$ to derive 
\begin{equation}
\frac{\partial g}{\partial v^i}=g\frac{q_i}{T_e}-\frac{m_e}{T_e}v_ig\, ,
\end{equation} 
and 
\begin{equation}
\label{gpp}
\int \frac{d^3q}{(2 \pi)^3} g q^i q^j =\delta^{ij} n_e m_e T_e+ n_e m_e^2 v^i v^j\, .
\end{equation}
Thus it is easy to compute
\begin{equation}
\label{finaleener}
e^{\Phi+\Psi}\frac{\Psi_{,k}}{m_e^3} \int \frac{d^3q}{(2 \pi)^3} g q^2q^k
= - e^{\Phi+\Psi} \Psi_{,k}v^2\frac{T_e}{m_e}+3e^{\Phi+\Psi} \Psi_{,k}v_kn_e\frac{T_e}{m_e}+
e^{\Phi+\Psi} \Psi_{,k}v_kv^2\, ,
\end{equation}
which is negligible taking into account that $T_e/m_e$ is of the order of a thermal velocity squared.  

With these results we are able to compute the left-hand side  of the Boltzmann equation~(\ref{boltzep}) integrated over 
$d^3q/(2\pi)^3$. The same operation must be done for the collision terms on the right hand side. For example for the 
first of the equations in~(\ref{boltzep}) this brings to the integrals 
$ \langle c_{ep} \rangle_{QQ'qq'} +\langle c_{e\gamma} \rangle_{pp'qq'}$. However looking at Eq.~(\ref{collisionterm}) one realizes 
that $\langle c_{e\gamma} \rangle_{pp'qq'}$ vanishes because the integrand is antisymmetric under the change ${\bf q} 
\leftrightarrow {\bf q'}$ and ${\bf p} 
\leftrightarrow {\bf p}'$. In fact this is simply a consequence of the fact that the electron number is conserved for this process. 
The same argument holds for the other term $\langle c_{ep} \rangle_{QQ'qq'}$. Therefore the right-hand side of 
Eq.~(\ref{boltzep}) integrated over $d^3q/(2\pi)^3$ vanishes and we can give the evolution equation for $n_e$. Collecting 
the results from Eq.~(\ref{r1}) to~(\ref{finaleener}) we find 
\begin{equation}
\frac{\partial n_e}{\partial \eta}+e^{\Phi+\Psi} \frac{\partial(v^i n_e)}{\partial x^i}+3(
{\cal H}-\Psi')n_e -2e^{\Phi+\Psi} \Psi_{,k}v^k+e^{\Phi+\Psi} \Phi_{,k}v^k=0\, .
\end{equation}
Similarly, for CDM particles, we find
\begin{equation}
\frac{\partial n_{\rm CDM}}{\partial \eta}+e^{\Phi+\Psi} \frac{\partial(v^i 
n_{\rm CDM})}{\partial x^i}+3(
{\cal H}-\Psi')n_{\rm CDM} -2 e^{\Phi+\Psi} \Psi_{,k}v_{\rm CDM}^k+e^{\Phi+\Psi} \Phi_{,k}v_{\rm CDM}^k=0\, .
\end{equation}

\subsection{Momentum continuity equation}
\noindent 
Let us now multiply Eq.~(\ref{Dg}) by $(q^i/E) /(2 \pi)^3$ and integrate over $d^3q$. In this way we will find the continuity 
equation for the momentum of baryons. The first term just gives $(n_e v^i)'$. The second integral is of the type 
\begin{equation}
\frac{\partial}{\partial x^j} \int \frac{d^3q}{(2 \pi)^3} g\, \frac{q n^j}{E} \frac{qn^i}{E} =
\frac{\partial}{\partial x^j}\left( n_e \frac{T_e}{m_e} \delta^{ij}+n_e v^i v^j \right)\, ,
\end{equation} 
where we have used Eq.~(\ref{gpp}) and $E=m_e$. The third term proportional to $({\cal H}-\Psi')$ is 
\begin{eqnarray}
\label{FirstI}
\int \frac{d^3q}{(2 \pi)^3} q^k \frac{\partial g}{\partial q_k} \frac{q^i}{E}=
4 n_e+\int \frac{d^3q}{(2 \pi)^3} g \frac{q^2}{E^2}\frac{q^i}{E}\, ,
\end{eqnarray}
where we have integrated by parts over $q^i$. Notice that the last term in Eq.~(\ref{FirstI}) is negligible being the same integral we 
discussewd above in Eq.~(\ref{finaleener}). By the same arguments that lead to neglect the term of Eq.~(\ref{finaleener}) it is easy to 
check that all the remaining integrals proportional to the gravitational potentials are negligible except for 
\begin{equation}
- e^{\Phi+\Psi} \Phi_{,k}\int \frac{d^3q}{(2 \pi)^3} \frac{\partial g}{\partial q_k} q^i=n_ee^{\Phi+\Psi}\Phi^{,i}\, . 
\end{equation}
The integrals proportional to the second-order vector and tensor perturbations vanish as 
vector and tensor perturbations are traceless and divergence-free. The 
only one which survives is the term proportional to $\omega^{i\prime}+{\cal H}\omega^{i}$ in Eq.~(\ref{Dg}). 

Therefore  for the integral over $d^3q q^i/E$ 
of the left-hand side of the Boltzmann equation~(\ref{Dg})
for a massive particle with mass $m_e$ ($m_p$) and distribution function~(\ref{gel}) we find 
\begin{eqnarray}   
\label{dgetau}   
\int \frac{d^3q}{(2\pi)^3} \frac{q^i}{E} 
\frac{d g_e}{d\eta} =  \frac{\partial (n_e v^i)}{\partial \eta}+4 ({\cal H}-\Psi') n_e v^i +\Phi^{,i} e^{\Phi+\Psi} 
n_e+ e^{\Phi+\Psi} \left( n_e \frac{T_e}{m_e} \right)^{,i}+ e^{\Phi+\Psi} \frac{\partial}{\partial x^j}(n_e v^j v^i) +
\frac{\partial \omega^i}{\partial \eta} n_e + {\cal H} \omega^i n_e \, . \nonumber \\
\end{eqnarray}

Now, in order to derive the momentum conservation equation for baryons, we take the first moment of both Eq.~(\ref{boltzep}) 
and~(\ref{boltzep2}) multiplying them by ${\bf q}$ and ${\bf Q}$ respectively and integrating over the momenta. Since previously we 
integrated the left-hand side of these equations over $d^3q q^i/E$, we just need to multiply 
the previous integrals by $m_e$ for the electrons and for $m_p$ for the protons. Therefore if we sum the 
first moment of Eqs.~(\ref{boltzep}) and~(\ref{boltzep2}) the dominant contribution on the left-hand side will be that of the protons
\begin{eqnarray}
\int \frac{d^3 Q}{(2 \pi)^3} Q^i\, \frac{dg_p}{d\eta}=\langle c_{ep} (q^i+Q^i)\rangle_{QQ'qq'} +\langle c_{e\gamma} q^i\rangle_{pp'qq'}\, .
\end{eqnarray}  
Notice that the integral of the Coulomb collision term $c_{ep} (q^i+Q^i)$ over all momenta vanishes simply because of momentum conservation 
(due to the Dirac function $\delta^4(q+Q-q'-Q')$). As far as the Compton scattering is concerned we have that, following 
Ref.~\cite{Dodelsonbook}, 
\begin{equation}
\langle c_{e\gamma} q^i \rangle_{pp'qq'} =- \langle c_{e\gamma} p^i \rangle_{pp'qq'} \, ,  
\end{equation}   
still because of the total momentum conservation. Therefore what we can compute now is the integral over all momenta of 
$c_{e\gamma} p^i$. Notice however that this is equivalent just to multiply the Compton collision term $C({\bf p})$ 
of Eq.~(\ref{collisionterm}) by $p^i$ and integrate over $d^3p/(2\pi^3)$ 
\begin{equation}
\label{Ci}
\langle c_{e\gamma} p^i \rangle_{pp'qq'} = \int \frac{d^3p}{(2\pi)^3} p^i C({\bf p})\, .
\end{equation}
where $C({\bf p})$ has been already computed in Eqs.~(\ref{C1p}) and~(\ref{C2p}). 

We will do the integral~(\ref{Ci}) in the following. First let us introduce the definition of the velocity of photons in terms of 
the distribution function 
\begin{equation}
\label{vp}
(\rho_\gamma+p_\gamma) v^i_\gamma = \int \frac{d^3 p}{(2 \pi)^3} f p^i\, ,
\end{equation}  
where $p_\gamma = \rho_\gamma/3$ is the photon pressure and $\rho_\gamma$ the energy density. At first-oder we get
\begin{equation}
\label{vp1}
\frac{4}{3} v^{(1) i}_\gamma= \int \frac{d\Omega}{4\pi} \Delta^{(1)}\, n^i \, ,
\end{equation}
where $\Delta$ is the photon distribution anisotropies defined in Eq.~(\ref{Delta2}). At second-order we instead find
\begin{equation}
\label{vp2}
\frac{4}{3} \frac{v^{(2) i}_\gamma}{2}= \frac{1}{2} \int \frac{d\Omega}{4\pi} \Delta^{(2)}\, n^i-\frac{4}{3} \delta^{(1)}_\gamma 
v^{(1)i}_\gamma \, .
\end{equation}
Therefore the terms in Eqs.~(\ref{C1p}) and~(\ref{C2p}) proportional to $f^{(1)}({\bf p})$ and $f^{(2)}({\bf p})$ 
will give rise to terms containing the velocity of the photons. On the other hand 
the terms proportional to $f^{(1)}_0(p)$ and $f^{(2)}_{00}(p)$, once integrated, vanish because of the integral over the momentum direction
$n^i$, $\int d\Omega n^i=0$. Also the integrals involving $P_2({\bf {\hat v}}\cdot {\bf n})=[3 ({\bf {\hat v}}\cdot {\bf n})^2-1]/2$ 
in the first line of Eq.~(\ref{C1p}) and~(\ref{C2p}) 
vanish since 
\begin{equation}
\int d\Omega P_2({\bf {\hat v}}\cdot {\bf n})\,  n^i= {\hat v}^k {\hat v}^j  \int d\Omega n_kn_jn^i=0\, ,
\end{equation}
where we are using the relations~(\ref{relOmega}). Similarly all the terms proportional to $v$, 
$({\bf v} \cdot {\bf n})^2$ and $v^2$ do not give any contribution to Eq.~(\ref{Ci}) and, 
in the second-order collision term, one can check that 
$\int d\Omega Y_2({\bf n}) n^i =0$. Then there are terms proportional to 
$({\bf v}\cdot {\bf n}) f^{(0)}(p)$, $({\bf v}\cdot {\bf n}) p \partial f^{(0)}/\partial p$ and 
$({\bf v}\cdot {\bf n}) p \partial f^{(1)}_0/\partial p$ for which we can use the rules~(\ref{rules}) when 
integrating over $p$ while the integration over the momentum direction is   
\begin{equation}
\int \frac{d\Omega}{4\pi} ({\bf v}\cdot {\bf n}) n^i =v_k  \int \frac{d\Omega}{4\pi}  n^kn^i= \frac{1}{3} v^i \, .
\end{equation}
Finally from the second line of Eq.~(\ref{C2p}) we get three integrals. One is 
\begin{equation}
\int \frac{d^3p}{(2\pi)^3} p^i \, ({\bf v}\cdot {\bf n}) f^{(1)}({\bf p})= \bar{\rho}_\gamma 
\int \frac{d\Omega}{4 \pi} \Delta^{(1)} ({\bf v}\cdot {\bf n})  n^i\, , 
\end{equation} 
where $\bar{\rho}_\gamma$ is the background energy density of the photons. 
The second comes from  
\begin{eqnarray}
\frac{1}{2} \int \frac{d^3p}{(2\pi)^3} p^i\,   ({\bf v} \cdot {\bf n}) P_2({\bf {\hat v}}\cdot {\bf n}) \left(f^{(1)}_2(p)-
p \frac{\partial f^{(1)}_2(p) }{\partial p} \right)&=& \frac{5}{4} \bar{\rho}_\gamma \Delta^{(1)}_2 
\left[ 3 
v_j  {\hat v}_k {\hat v}_l \int \frac{d\Omega}{4 \pi} n^in^jn^kn^l-v_j \int \frac{d\Omega}{4 \pi} n^in^j 
\right] \nonumber \\
&=& \frac{1}{3} \bar{\rho}_\gamma \Delta^{(1)}_2 {\hat v}^i\, ,
\end{eqnarray}
where we have used the rules~(\ref{rules}), Eq.~(\ref{relOmega}) and $\int (d\Omega/4 \pi)\, n^in^jn^kn^l = (\delta^{ij} \delta^{kl} 
+\delta^{ik} \delta^{lj}+\delta^{il} \delta^{jk})/15$. In fact the third integral 
\begin{equation}
- \int \frac{d^3p}{(2\pi)^3} p^i ({\bf v} \cdot {\bf n}) f^{(1)}_2(p)\, , 
\end{equation}
exactly cancels the previous one. Summing the various integrals we find  
\begin{equation}
\label{Ciint}
\int\frac{d{\bf p}}{(2\pi)^3} C({\bf p}) {\bf p}= 
n_e \sigma_T {\bar \rho_\gamma} \Bigg[ \frac{4}{3} ({\bf v}^{(1)}-{\bf v}_\gamma^{(1)})
-\int \frac{d\Omega}{4\pi} \frac{\Delta^{(2)}}{2} {\bf n}+ \frac{4}{3} \frac{{\bf v}^{(2)}}{2}+
\frac{4}{3} \delta^{(1)}_e ({\bf v}^{(1)}-{\bf v}_\gamma^{(1)})+ \int \frac{d\Omega}{4\pi} \Delta^{(1)} ({\bf v} \cdot {\bf n}) {\bf n} 
+ \Delta_0^{(1)} {\bf v} \Bigg]. \\
\end{equation}
Eq.~(\ref{Ciint}) can be further simplified. Recalling that $\delta^{(1)}_\gamma = \Delta^{(1)}_0$ we use Eq.~(\ref{vp2}) and notice that 
\begin{equation}
\int \frac{d\Omega}{4\pi} \Delta^{(1)}\, ({\bf v} \cdot {\bf n}) n^i = v_j^{(1)} \Pi^{ji}_\gamma+\frac{1}{3} v^i \Delta^{(1)}_0 \, ,
\end{equation}
where $\Pi^{ij}_\gamma$ is the quantity defined in Eq.~(\ref{quadrupole}). 

Thus our final expression for the integrated collision term~(\ref{Ci}) reads
\begin{equation}
\label{Cifinal}
\int \frac{d^3 p}{(2\pi)^3} C({\bf p}) p^i =
n_e \sigma_T {\bar \rho_\gamma}
\left[ \frac{4}{3} (v^{(1)i}-v_\gamma^{(1)i}) + \frac{4}{3} \left( \frac{v^{(2)i}}{2}-\frac{v_\gamma^{(2)i}}{2} \right) +
\frac{4}{3} \delta^{(1)}_e (v^{(1)i}-v_\gamma^{(1)i})+v^{(1)}_j \Pi^{ji}_\gamma \right ]\, .
\end{equation}

We are now able to give the momentum continuity equation by for baryons by combining $m_p dg_p/d\eta$ from Eq.~(\ref{dgetau}) with the 
collision term~(\ref{Cifinal})
\begin{eqnarray}
\label{mcc}
& & \frac{\partial (\rho_b v^i)}{\partial \eta}+4 ({\cal H}-\Psi') \rho_b v^i +\Phi^{,i} e^{\Phi+\Psi} 
\rho_b+ e^{\Phi+\Psi} \left( \rho_b \frac{T_b}{m_p}\right)^{,i}+ e^{\Phi+\Psi} \frac{\partial}{\partial x^j}(\rho_b v^j v^i) +
\frac{\partial \omega^i}{\partial \eta} \rho_b+ {\cal H} \omega^i \rho_b  \nonumber \\
& & = - n_e \sigma_T a\, {\bar \rho_\gamma}
\left[ \frac{4}{3} (v^{(1)i}-v_\gamma^{(1)i}) + \frac{4}{3} \left( \frac{v^{(2)i}}{2}-\frac{v_\gamma^{(2)i}}{2} \right) +
\frac{4}{3} \delta^{(1)}_b (v^{(1)i}-v_\gamma^{(1)i})+v^{(1)}_j \Pi^{ji}_\gamma \right ] \, ,
\end{eqnarray}
where $\rho_b$ is the baryon energy density and, as we previously explained, 
we took into account that to a good 
approximation the electrons do not contribute to the mass of baryons. 
In the following we will expand explicitly at first and second-order Eq.~(\ref{mcc}).
\\
\\
{\bf First-order momentum continuity equation for baryons}
\\
\\
At first order we find
\begin{equation}
\label{mcc1}
\frac{\partial v^{(1)i}}{\partial \eta} +{\cal H} v^{(1)i}+\Phi^{(1),i}=\frac{4}{3} \tau' \frac{{\bar \rho}_\gamma}{{\bar \rho}_b} 
\left(  v^{(1)i}-v^{(1)i}_\gamma \right)\, .
\end{equation} 
\\
\\
{\bf Second-order momentum continuity equation for baryons}
\\
\\
At second-order there are various simplifications. In particular notice that the term on the right-hand side of Eq.~(\ref{mcc}) which is proportional 
to $\delta_b$ vanishes when matched to expansion of the left-hand side by virtue of the first-order equation~(\ref{mcc1}). Thus at the end 
we find a very simple equation
\begin{eqnarray}
& & \frac{1}{2} \left( \frac{\partial v^{(2)i}}{\partial \eta} +{\cal H} v^{(2)i} + 
2 \frac{\partial \omega^i}{\partial \eta} +2 {\cal H} \omega_i  + \Phi^{(2),i}\right) - \frac{\partial \Psi^{(1)}}{\partial \eta} v^{(1)i}
+ v^{(1)j}\partial_jv^{(1)i}+(\Phi^{(1)}+\Psi^{(1)}) \Phi^{(1),i} +\left( \frac{T_b}{m_p} \right)^{,i} \nonumber \\
& & = \frac{4}{3} \tau ' \frac{{\bar \rho}_\gamma}{{\bar \rho}_b} \left[ 
\left(  \frac{v^{(2)i}}{2}-\frac{v^{(2)i}_\gamma}{2}  \right) +\frac{3}{4} v^{(1)}_j \Pi^{ji}_\gamma 
\right] \, ,
\end{eqnarray}   
with $\tau'=- {\bar n}_e \sigma_T a$.
\\
\\
{\bf First-order momentum continuity equation for CDM}
\\
\\
Since CDM particles are collisionless, at first order we find
\begin{equation}
\label{mcc1}
\frac{\partial v_{\rm CDM}^{(1)i}}{\partial \eta} +{\cal H} 
v_{\rm CDM}^{(1)i}+\Phi^{(1),i}=0\, .
\end{equation} 
\\
\\
{\bf Second-order momentum continuity equation for CDM}
\\
\\
At second-order we find 
\begin{equation}
\label{mcc2}
\frac{1}{2} \left( \frac{\partial v_{\rm CDM}^{(2)i}}{\partial \eta} +{\cal H} v_{\rm CDM}^{(2)i} + 
2 \frac{\partial \omega^i}{\partial \eta} +2 {\cal H} \omega_i  + \Phi^{(2),i}\right) - \frac{\partial \Psi^{(1)}}{\partial \eta} v_{\rm CDM}^{(1)i}
+ v_{\rm CDM}^{(1)j}\,  \partial_j v_{\rm CDM}^{(1)i}+(\Phi^{(1)}+\Psi^{(1)}) \Phi^{(1),i} 
+\left( \frac{T_{\rm CDM}}{m_{\rm CDM}} \right)^{,i} 
=0\, .
\end{equation}   
\section{Energy momentum tensors}
\noindent 
In this section we provide the expressions for the energy-momentum tensors for photons and massive particles in terms of their 
distribution fucntions. 
\subsection{Energy momentum tensor for photons}
\noindent 
The energy momentum tensor for photons is defined as

\begin{equation}
T^\mu_{\gamma ~\nu}=\frac{2}{\sqrt{-g}}\int \frac{d^3 P}{(2\pi)^3}\, 
\frac{P^\mu P_\nu}{P^0}\, f\, ,
\end{equation}
where $g$ is the determinant of the metric (\ref{metric}) and $f$
is the distribution function. We thus obtain

\begin{eqnarray}
\label{T00photons}
T^0_{\gamma ~0}&=&-\bar{\rho}_\gamma
\left(1+\Delta_{00}^{(1)}+\frac{\Delta_{00}^{(2)}}{2}\right)\, ,\\
\label{Ti0photons}
T^i_{\gamma ~0}&=&-\frac{4}{3}e^{\Psi+\Phi}\bar{\rho}_\gamma\left(
v_\gamma^{(1)i}+
\frac{1}{2}v_\gamma^{(2)i}+\Delta^{(1)}_{00} v_\gamma^{(1)i}\right)+\frac{1}{3}
\bar{\rho}_\gamma e^{\Psi-\Phi}\omega^i\, ,\\
\label{Tijphotons}
T^i_{\gamma ~j}&=& \bar{\rho}_\gamma\left(\Pi^i_{\gamma ~j}+
\frac{1}{3}\delta^i_{~j}\left(1+\Delta_{00}^{(1)}+\frac{\Delta_{00}^{(2)}}{2}\right)
\right)
\, ,
\end{eqnarray}
where $\bar{\rho}_\gamma$ is the background energy density of photons
and 

\begin{equation}
\label{quadrupole}
\Pi^{ij}_{\gamma}=\int\frac{d\Omega}{4\pi}\,\left(n^i n^j-\frac{1}{3}
\delta^{ij}\right)\left(\Delta^{(1)}+\frac{\Delta^{(2)}}{2}\right)\, ,
\end{equation}
are the quadrupole moments of the photons. The photon velocity is defined in Eqs.~(\ref{vp}),~(\ref{vp1}) and~(\ref{vp2}).

\subsection{Energy momentum tensor for massive particles}
\noindent 
The energy momentum tensor for massive particles of mass $m$, number
density $n$ and degrees of freedom $g_d$ is 

\begin{equation}
T^\mu_{m ~\nu}=\frac{g_d}{\sqrt{-g}}\int \frac{d^3 Q}{(2\pi)^3}\, 
\frac{Q^\mu Q_\nu}{Q^0}\, g(q^i,x^\mu,\eta)\, ,
\end{equation}
where $g$ is the distribution function. For electrons (or protons) $g_d=2$ (we are not counting antiparticles). 
We obtain

\begin{eqnarray}
\label{T00massive}
T^0_{m~0}&=&-\rho_m=-\bar{\rho}_m\left(1+\delta^{(1)}_m+\frac{1}{2}\delta^{(2)}
\right)\, ,\\
\label{Ti0massive}
T^i_{m~0}&=&-e^{\Psi+\Phi}\rho_m v_m^{i}=
-e^{\Psi+\Phi}\bar{\rho}_m\left(v_m^{(1)i}+
\frac{1}{2}v_m^{(2)i}+\delta^{(1)}_m v_m^{(1)i}\right)\, ,
\\
\label{Tijmassive}
T^i_{m~j}&=& \rho_m\, \left(\delta^i_{~ j} \frac{T_m}{m}+v_m^{i} v_{m~j}\right)=
\bar{\rho}_m\left(\delta^i_{~ j} \frac{T_m}{m}+v_m^{(1)i} v^{(1)}_{m~j}\right)
\, .
\end{eqnarray}
where $\bar{\rho}_m$ is the background energy density of the massive
particles and we have included the equilibrium temperature $T_m$. The velocity are defined in terms of the distribution function 
in Eq.~(\ref{defvg}).

\section{Summary}
\begin{table}[ht]
\centering
\begin{tabular}{c c c c c c}
\hline\hline 
\vspace{0.2cm}
Symbol & & & & Definition & Equation \\
\hline  
$\Phi,\Psi$ & & & & Gravitational potentials in Poisson gauge & 
(\ref{metric}) \\
$\omega_i$ & & & & $2$nd-order vector perturbation in Poisson gauge & 
(\ref{metric}) \\
$\chi_{ij}$ & & & & $2$nd-order tensor perturbation in Poisson gauge & 
(\ref{metric}) \\
$\eta$ & & & & Conformal time & (\ref{metric})  \\
$f$ & & & & Photon distribution function & (\ref{Df}) \\
$g$ & & & & Distribution function for massive particles & (\ref{gel}) \& (\ref{DG}) \\
$f^{(i)}$ & & & & $i$-th order perturbation of the photon distribution function  & (\ref{expf}) \\
$f^{(i)}_{\ell m}$ & & & & Moments of the photon distribution function & (\ref{angular}) \\
$C({\bf p})$ & & & & Collision term & (\ref{collisionterm}) \& (\ref{Integralcolli}) \\ 
$p$ & & & & Magnitude of photon momentum (${\bf p}=pn^i$) &  (\ref{defp}) \\
$n^i$ & & & & Propagation direction & (\ref{Pi}) \\
$\Delta^{(1)}(x^i,n^i,\eta)$ & & & & First-order fractional energy photon fluctuations & (\ref{Delta1})  \\
$\Delta^{(2)}(x^i,n^i,\eta)$ & & & & Second-order fractional energy photon fluctuations & (\ref{Delta2}) \\
$n_e$ & & & & Electron number density & (\ref{defne}) \\
$\delta_e (\delta_b)$ & & & & Electron (baryon) density perturbation & (\ref{deltac1}) \\
${\bf k}$ & & & & Wavenumber & (\ref{BF}) \\
$v_m$ & & & & Baryon velocity perturbation & (\ref{vzero}) \& (\ref{vdec})  \\ 
$v^{(2)i}_{\rm CDM}$ & & & & Cold dark matter velocity & (\ref{mcc2}) \\
$v^{(2)i}_\gamma$ & & & & Second-order photon velocity & (\ref{vp2}) \\
$S_{\ell m}$ & & & & Temperature source term & (\ref{H}) \\  
$\tau$ & & & & Optical depth & (\ref{deftau}) \\
${\bar \rho}_\gamma ({\bar \rho}_b)$ & & & & Background photon (baryons) energy density & (\ref{mcc1}) \\ 
\hline\hline
\end{tabular}
\end{table}

\noindent
In this paper we took the first step towards the evaluation of the
full radiation transfer at second order in perturbation theory
by computating  the second-order 
Boltzmann equations 
describing the evolution of the baryon-photon fluid. They
allow to follow the time evolution of the CMB anisotropies at second-order
and at all scales
from the early epoch,  when the cosmological perturbations were generated,
to the present  through the recombination era. The dynamics at the second order
is particularly important when dealing with the issue
of NG in the CMB anisotropies: many mechanisms for the generation of the
primordial inhomogeneities  predict a level of NG in the 
curvature perturbation which might
be detectable by present and future experiments. Having an accurate 
theoretical prediction of the CMB anisotropy NG in terms of the 
 primordial non-Gaussian 
seeds  is somewhat mandatory. 
This paper will be followed by  a companion one where 
we will present the computation of the three-point correlation
function at recombination making use of the set of Boltzmann 
equations derived here. 

\vskip 0.5cm 
\centerline{\bf Acknowledgments}
\vskip 0.5cm
\noindent
A.R. is on 
leave of absence from INFN, Padova. N.B. is partially supported by INFN.
\newpage 

\end{document}